\documentclass[%
 reprint,
 amsmath,amssymb,
 aps,
 prapplied,
 floatfix,
 nobibnotes,
 nofootinbib,
]{revtex4-2}

\usepackage{graphicx}
\usepackage{dcolumn}
\usepackage{bm}
\usepackage{xurl}
\usepackage{hyperref}

\makeatletter
\@booleanfalse\footinbib@sw
\let\frontmatter@footnote@produce\frontmatter@footnote@produce@footnote
\def\@fnsymbol#1{\ifcase#1\or *\or \dagger\or \ddagger\or \mathsection\or \mathparagraph\or \|\or **\or \dagger\dagger\or \ddagger\ddagger \else\@ctrerr\fi}
\makeatother

\begin{document}

\title{Adjoint-State Identifiability of Piezo-Tunable Valley Splitting in 2D Magnetic Heterostructures}

\author{Suhas Suresh Bharadwaj}
\email{f20230029@dubai.bits-pilani.ac.in}
\affiliation{Department of Electrical and Electronics Engineering, BITS Pilani, Dubai Campus, Dubai, United Arab Emirates}
\date{\today}

\begin{abstract}
Mechanical strain has been proposed as a route to dynamically tune proximity-induced valley splitting in two-dimensional magnetic heterostructures for solid-state information processing. While first-principles studies routinely predict strain-modulated valley splittings at the atomic scale, they rarely address whether the underlying coupling parameter can be extracted from a macroscopic transport measurement against realistic device noise. This paper develops a classical, partial differential equation-constrained multiscale inverse framework to quantify the device-level identifiability of the exchange-strain coupling from a transverse Hall voltage in a molybdenum disulfide and chromium tribromide heterostructure. Structural relaxation confirms a chiral $C_3$ point-group symmetry of the Hollow-stacked structure, and Neumann's principle reduces the coupling tensor to a single scalar $\eta$. Density functional theory with a dispersion correction gives a zero-strain splitting of $-0.45$ meV and bounds the coupling to $|\eta| \lesssim 2$ meV at small tensile strain while leaving its sign open, motivating the transport analysis. For an example coupling of $|\eta| = 0.56$ meV, where an actuator strain trace of $3.5 \times 10^{-4}$ shifts the valley splitting by only $0.2~\mu$eV, operation at 10 K below the Curie temperature of $\text{CrBr}_3$ makes the coupling statistically resolvable from 5.6 to 3,704 V/cm. Even under a demanding stress test in which $\eta$ is recovered jointly with the zero-strain splitting, an uncalibrated Hall gain, and an unknown intervalley scattering time, the window spans 25 to 3,704 V/cm and the Cram\'er-Rao bound at the 10\% velocity-saturation limit reaches $13.6~\mu$eV, 21 times below the resolution criterion. Since this window covers the full operational bias range of 2D field-effect transistors, from low-power linear readout to high-field drive, this framework serves as a practical pre-fabrication diagnostic for valleytronic devices.
\end{abstract}

\maketitle

\section{Introduction}

The valley degree of freedom in two-dimensional transition metal dichalcogenides provides a platform for solid-state information processing \cite{xiao2012coupled, mak2014valley, xu2014spin, schaibley2016valleytronics, rycerz2007valley}, provided the intrinsic valley degeneracy can be lifted in a controlled and tunable manner \cite{macneill2015breaking}. In non-magnetic monolayers, time-reversal symmetry protects this degeneracy and forbids mechanical strain alone from inducing valley splitting in the pristine material \cite{rostami2015theory, amorim2016novel}; as formalized in Proposition 1 (Sec. II.B), a static lattice deformation is even under time reversal and therefore cannot lift the degeneracy of quantum states linked by that symmetry. Overcoming this constraint traditionally requires strong external magnetic fields, which limits compact device integration.

An alternative mechanism is the magnetic proximity effect, in which a non-magnetic transition metal dichalcogenide is electronically coupled to an adjacent magnetic substrate so that interfacial exchange breaks time-reversal symmetry and produces a static valley splitting without an external field \cite{zhao2017enhanced, zhong2017van, zutic2019proximity, scharf2017magnetic, norden2019giant, ciorciaro2020observation}. While this static response provides the required symmetry breaking, it lacks the continuous tunability that active analog device operation needs. To introduce dynamic control, recent theoretical work has turned to mechanical strain modulation, almost entirely in intrinsically magnetic two-dimensional materials such as altermagnets \cite{smejkal2022giant}. Density functional theory calculations frequently predict significant piezovalley effects in altermagnetic Janus monolayers such as $\text{Cr}_2\text{SSe}$ \cite{shen2026strain} and $\text{V}_2\text{STeO}$ \cite{chang2025valley}, in ferrovalley candidates such as $\text{VSSe}$ \cite{nie2019first} and $\text{VSeS}$ \cite{li2022controlled}, and in altermagnetic iron-based chalcogenides \cite{fan2025valley}, although restricting attention to intrinsically magnetic monolayers narrows the choice of synthesizable materials compared with proximity-coupled heterostructures.

More importantly, the first-principles piezovalley studies cited above leave a methodological gap because they work only in the forward direction at a single microscopic length scale. In this literature, a strain-induced energy splitting is predicted from the microscopic band structure and the analysis stops there, without connecting the predicted splitting to a macroscopic device observable or evaluating whether the underlying coupling can be statistically recovered against realistic background noise. Exchange-strain coupling constants are routinely calculated as deterministic scalar values rather than unknown parameters to be extracted from transport measurements. Because microscopic material predictions are generated far faster than experiments can fabricate and test them, a general diagnostic is needed to determine which predicted couplings are actually resolvable in a working two-dimensional transistor or Hall-bar device before fabrication.

Bridging this gap matters because every signal in the physical chain is small. For an example coupling of 0.56 meV and an actuator strain trace of $3.5 \times 10^{-4}$, the valley splitting changes by only 0.2 $\mu$eV, and the part modulated by the actuator voltages is 0.04 $\mu$eV. Even a $+1$\% strain sweep in density functional theory shifts the splitting by only about 10 $\mu$eV, which sits right at the $\pm 15~\mu$eV numerical precision set by the cutoff, basis-set size, and $k$-mesh in our 20-atom supercell calculations. In other words, first-principles theory can bound the magnitude of the exchange-strain coupling, but it cannot resolve its exact value or sign, and this scale contrast motivates a device-level inverse framework. Because the coupling must be read out through a transverse Hall voltage against thermal and amplifier noise, whether a device can recover it is a question of identifiability that depends on the field, the geometry, and the noise as much as on the material. This paper treats the exchange-strain coupling as an unknown parameter of a multiscale model and uses an adjoint-state inverse framework to compute the voltage sensitivity and the Fisher information bounds on its recovery in a molybdenum disulfide ($\text{MoS}_2$) monolayer on a chromium tribromide ($\text{CrBr}_3$) substrate \cite{ciorciaro2020observation}.

This paper addresses this gap with three results. First, structural relaxation of the $\text{MoS}_2/\text{CrBr}_3$ van der Waals heterostructure shows that the Hollow and Chromium-aligned stackings have a chiral $C_3$ symmetry (space group P3), whereas the Bromine-aligned registry has no symmetry (space group P1), lowering the idealized $C_{3v}$ symmetry commonly assumed in effective-Hamiltonian studies on the basis of unrelaxed lattice alignments \cite{scharf2017magnetic, zutic2019proximity}. A formal Neumann's-principle proof shows that the scalar reduction $\eta_{ij} = \eta\delta_{ij}$ survives this symmetry lowering because the antisymmetric tensor part allowed by the three-fold axis is orthogonal to the symmetric strain tensor, whereas the Bromine-aligned registry retains independent in-plane components whose Fisher information matrix is singular under any single fixed-ratio strain sweep. In the ground-state Hollow registry, including the dispersion correction raises the zero-strain splitting by a factor of 75 to $-0.45$ meV and bounds the coupling to $|\eta| \lesssim 2$ meV at small tensile strain without fixing its sign.

Second, a classical partial differential equation-constrained adjoint-state inverse-problem framework is developed \cite{lions1971optimal, tarantola2005inverse, nocedal2006numerical} to recover $\eta$ directly from spatially resolved transverse voltage measurements governed by the transport equations. The full continuous adjoint formulation is derived for the valley-resolved drift-diffusion-Poisson system \cite{markowich1986stationary, gajewski1985existence}, and Fisher information bounds are computed from its linear drift-diffusion reduction at a prescribed longitudinal field \cite{cramer1946mathematical, rao1945information}.

Third, the field conditions required for statistical resolvability are determined by combining the Caughey-Thomas high-field mobility model \cite{caughey1967carrier, smithe2018high} with the strain trace from the piezoelectric actuator model. At 10 K, below the Curie temperature of $\text{CrBr}_3$, an example coupling of 0.56 meV is resolvable within an operating window from 5.6 to 3,704 V/cm whose lower bound scales as $1/|\eta|$. Thermal and amplifier noise set the lower bound, while the conservative 10 percent velocity-saturation limit ($\approx 30$ V across the channel) sets the upper bound, with the 20 and 40 percent limits at 8,333 V/cm and 22.2 kV/cm serving as extended high-field references. Recovering $\eta$ jointly with the zero-strain splitting, the Hall gain, and the intervalley scattering time raises the lower bound to 25 V/cm, while for a substrate that remains ordered at 300 K the window spans 860 to 3,704 V/cm. Because this nearly three-decade window covers the full operational bias scale of two-dimensional field-effect transistors, from low-power linear channel readout at tens of millivolts to high-field on-state drive near velocity saturation, the framework establishes where a predicted coupling can be resolved under standard device operating conditions.

The rest of the paper follows the order of these results. Section II sets up the multiscale forward model, from first-principles calculations to continuum electromechanical actuation, and Section III derives the continuous adjoint formulation and the statistical identifiability analysis \cite{kiefer1959optimum}. Section IV presents the density functional theory parameterization, the device-level identifiability window, its temperature dependence, and the joint recovery of the coupling with the zero-strain splitting, the Hall gain, and the intervalley scattering time. Section V discusses the scope of the central claim and the approximations of the model chain, and Section VI concludes with the broader applications of this inverse framework across topological and magnetic two-dimensional materials.

\begin{table*}[htbp]
\caption{\label{tab:model_parameters} Physical parameters, symbols, and nominal values used in the multiscale forward model.}
\centering
\renewcommand{\arraystretch}{1.3}
\begin{tabular}{llc}
\\
\hline\hline
\textbf{Symbol} & \textbf{Physical Representation} & \textbf{Nominal Value} \\
\hline
\multicolumn{3}{c}{\textit{Microscopic \& Band Structure Parameters}} \\
\hline
$a$ & Lattice constant & $3.184$ \AA \\
$\Delta$ & Bare direct band gap & $1.678$ eV \\
$t$ & Effective hopping parameter & $1.1601$ eV \\
$m^*$ & Effective mass at the $K$-valley minimum & $0.4686 m_e$ \\
$\lambda_{SO}$ & Valence-band spin-orbit splitting & $147.8$ meV \\
$M_0$ & Zero-strain valley splitting (D3) & $-0.452$ meV \\
$\eta$ & Exchange-strain coupling constant & $|\eta| \lesssim 2$ meV, $0.56$ meV used \\
$\beta$ & Leading-order Berry curvature parameter & -- \\
\hline
\multicolumn{3}{c}{\textit{Continuum \& Piezoelectric Parameters}} \\
\hline
$\varepsilon_{ij}$ & In-plane mechanical strain tensor & -- \\
$C_{ijkl}$ & Elasticity tensor (Aluminum Nitride) & (Detailed in SI) \\
$e_{kij}$ & Piezoelectric coupling tensor (Aluminum Nitride) & (Detailed in SI) \\
$\kappa_{ij}$ & Permittivity tensor (relative value $9$, isotropic) & $9\epsilon_0$ \\
$\nu$ & Strain ratio $\varepsilon_{yy}/\varepsilon_{xx} = -\nu$ used in the symmetry-collapse illustration & $0.27$ \\
\hline
\multicolumn{3}{c}{\textit{Device \& Transport Parameters}} \\
\hline
$n_K, n_{K'}$ & Valley-resolved two-dimensional carrier densities & -- \\
$\phi$ & Macroscopic electrostatic potential & -- \\
$J_\tau$ & Valley-resolved carrier flux & -- \\
$\tau_v$ & Intervalley scattering time & $1$ ns \\
$v_{sat}$ & Carrier saturation velocity limit & $2.0 \times 10^6$ cm/s \\
$T$ & Background operational temperature & $10$ K \\
$v_n$ & Amplifier noise floor limit & $5 \text{ nV}/\sqrt{\text{Hz}}$ \\
$d_{thickness}$ & Effective two-dimensional electrical thickness & $0.65$ nm \\
$\mu_0$ & Low-field mobility & $60$ cm$^2$/V\,s \\
$n_c$ & Contact carrier density per valley & $10^{18}$ cm$^{-3}$ \\
$B$ & Detection bandwidth & $1$ Hz \\
$N_s$ & Sensor readings treated as independent & $200$ \\
$\ell_p$ & Hall-probe length, equal to the sensor pitch & $0.4$ $\mu$m \\
$L$, $W$ & Channel length and width & $80$, $10$ $\mu$m \\
\hline\hline
\end{tabular}
\end{table*}

\section{Multiscale Forward Model}

The model connects first-principles electronic structure to continuum elastodynamics and macroscopic device transport, so that a microscopic atomic distortion can be related to a measurable voltage.

\subsection{Microscopic Valley Hamiltonian}

The model starts at the atomic level with the two-band $\mathbf{k} \cdot \mathbf{p}$ valley Hamiltonian \cite{xiao2012coupled}. Near the $K$ and $K'$ valleys, the unperturbed Hamiltonian is
\begin{equation}
H_0(\tau, \mathbf{k}) = a t (\tau k_x \sigma_x + k_y \sigma_y) + \frac{\Delta}{2} \sigma_z - \tau \frac{\lambda_{SO}}{2} (\sigma_z - 1) s_z
\end{equation}
where $\tau = \pm 1$ is the valley index and $s_z$ is the physical spin. The Pauli matrices $\sigma$ act in the basis of the conduction and valence orbitals. The four parameters are the lattice constant $a$, the effective hopping parameter $t$, the bare direct band gap $\Delta$, and the valence-band spin-orbit splitting $\lambda_{SO}$. They are fixed from first-principles calculations and are not fitting variables. All model parameters are listed in Table~\ref{tab:model_parameters}. The corresponding Berry curvature is
\begin{equation}
\Omega_{\tau,s}(\mathbf{k}) = -\tau \frac{2 a^2 t^2 \Delta_s}{(\Delta_s^2 + 4 a^2 t^2 k^2)^{3/2}}.
\end{equation}
Here $\Delta_s = \Delta - \tau s \lambda_{SO}$ is the gap for spin $s = \pm 1$ in valley $\tau$. This Berry curvature is odd in $\tau$. The asymmetry drives the macroscopic anomalous valley Hall observable derived later in this section.

\subsection{Symmetry Constraints and Magnetic Proximity}

Mechanical deformation of a pristine non-magnetic material does not produce a resolvable valley splitting. Proposition 1 states this constraint.

\textbf{Proposition 1.} Let $H(\varepsilon)$ be any Hamiltonian built from the strain tensor $\varepsilon_{ij}$ to arbitrary order acting on a time-reversal-symmetric non-magnetic crystal. Then $E_{K,s}(\varepsilon) = E_{K',-s}(\varepsilon)$ for every physical spin $s$ and every $\varepsilon$. Because strong intrinsic spin-orbit coupling locks a definite spin species to each valley at the relevant band edge, the states actually populated and probed at $K$ and $K'$ remain exactly time-reversal partners, and hence exactly degenerate, under any strain.

\emph{Proof.} The time-reversal operator is $\Theta = i s_y \mathcal{K}$, where $\mathcal{K}$ is complex conjugation, and it is antiunitary. A static strain $\varepsilon_{ij}$ is a real tensor that is even under time reversal, so for a non-magnetic crystal $\Theta H(\varepsilon) \Theta^{-1} = H(\varepsilon)$ at every order in $\varepsilon$. Time reversal maps a Bloch state with wavevector $\mathbf{k}$ and spin $s$ onto one with wavevector $-\mathbf{k}$ and spin $-s$ at the same energy, so $E_n(\mathbf{k}, s; \varepsilon) = E_n(-\mathbf{k}, -s; \varepsilon)$. The $K'$ valley sits at $-K$ (up to a reciprocal lattice vector), which gives $E_{K,s}(\varepsilon) = E_{K',-s}(\varepsilon)$ for every $\varepsilon$. Strong spin-valley locking pins one spin species to each valley at the band edge, so the populated states at $K$ and $K'$ are exactly these partners \cite{rostami2015theory, amorim2016novel}. In plain terms, strain deforms the lattice in the same way for both valleys and cannot tell $K$ from $K'$ unless something else breaks time-reversal symmetry. The heterostructure includes a magnetic proximity substrate for this reason, instead of applying strain directly to pristine non-magnetic monolayers \cite{scharf2017magnetic, zutic2019proximity}.

\subsection{Materials System and First-Principles Parameterization}

The system is a molybdenum disulfide ($\text{MoS}_2$) monolayer on a chromium tribromide ($\text{CrBr}_3$) magnetic substrate. A commensurate supercell of a $2 \times 2$ $\text{MoS}_2$ lattice and a $1 \times 1$ $\text{CrBr}_3$ lattice is used. The relaxed lattice constant of isolated CrBr$_3$ is 6.436~\AA\ in the same PBE setup, so the supercell constant of 6.368~\AA\ compresses the CrBr$_3$ layer by 1.06\%. First-principles relaxations of three stacking registries (Hollow, Chromium-aligned, and Bromine-aligned) are done with PBE and with PBE plus the DFT-D3 dispersion correction. All runs are collinear, with the total moment fixed at 6.00 $\mu_B$, a $\Gamma$-centered $6 \times 6 \times 1$ mesh, and a force threshold of 5 meV/\AA. With PBE alone, the total energies agree to within about 1 meV per supercell, so they do not single out one registry. With D3, the Chromium-aligned and Bromine-aligned registries lie 5.04 and 13.08 meV above Hollow. Hollow stacking is used as the reference structure. In the Hollow structure, the vertical gap between the lowest Br plane and the highest S plane is 4.26~\AA\ with PBE and 3.32~\AA\ with D3.

The calculations use VASP 6.5.1 with PAW potentials and the PBE functional, with no Hubbard $U$. The D3 calculations add the zero-damping DFT-D3 correction (IVDW = 11) at the PBE in-plane lattice constant. The spin-orbit calculations use a plane-wave cutoff of 600 eV, Gaussian smearing with a width of 0.05 eV, and an electronic convergence threshold of $10^{-7}$ eV. Spin-orbit coupling is included in the noncollinear calculations, with the spin axis along the out-of-plane direction and an initial moment of $3~\mu_B$ on each Cr atom. The cell height along $z$ is 35~\AA.

Structural relaxation followed by automated symmetry analysis, with the tolerance swept from $10^{-6}$ to $10^{-2}$, confirms that the relaxed Hollow structure has a chiral $C_3$ point-group symmetry, space group P3, with PBE and with D3. The Chromium-aligned structure also has space group P3, and the Bromine-aligned structure has no symmetry (space group P1). This lowers the originally hypothesized $C_{3v}$ symmetry. The density functional theory calculations on the isolated monolayer give a lattice constant of $a = 3.184$ \AA, a direct gap of $\Delta = 1.678$ eV, a valence-band spin-orbit splitting of $147.8$ meV, and a conduction-band spin-orbit splitting of $3.2$ meV. A parabolic fit restricted to the $K$-valley conduction band minimum gives an effective mass of $m^* = 0.4686 m_e$ and an effective hopping parameter of $t = 1.1601$ eV. These values are consistent with independent literature estimates \cite{xiao2012coupled, kormanyos2015kp}.

With the D3 correction, the magnetic substrate induces a zero-strain valley splitting of $M_0 = -0.452$ meV. This is the signed average of the $K$--$K'$ energy differences of the two spin partners of the $\text{MoS}_2$ conduction-band doublet in the relaxed 0\% structure, defined in Sec. II.D. With PBE alone, at the larger interlayer gap, $M_0 = -0.006$ meV. An orbital projection isolates the $\text{MoS}_2$-derived states from the substrate states, because the Type-II band alignment places substrate states close in energy to the target bands.

The D3 value is about six times smaller than the exciton valley splitting of 2.9 meV measured for MoSe$_2$/CrBr$_3$ \cite{ciorciaro2020observation}, and a splitting of similar size has been reported for WSe$_2$/CrI$_3$ \cite{zhong2017van}. The two quantities differ, since an exciton splitting includes the shifts of both band edges and the calculated value is that of the conduction band alone. The calculated splitting also depends on the stacking registry and the interlayer separation, as the present calculations show. With D3, $M_0$ is $-0.248$ and $-0.2104$ meV for the Chromium-aligned and Bromine-aligned registries, and the D3 correction changes it by a factor of 75 relative to PBE.

\subsection{Neumann's Principle and the Exchange-Strain Coupling}

The device relies on dynamic modulation of this proximity splitting by piezoelectric strain. The exchange-strain coupling is defined as
\begin{equation}
M(\varepsilon) = M_0 + \eta_{ij} \varepsilon_{ij}
\end{equation}
Neumann's principle \cite{nye1985physical} reduces the number of independent components of this coupling tensor. For any point group containing a three-fold rotational axis, the symmetric part of the in-plane coupling tensor must be isotropic, so that
\begin{equation}
\eta_{ij} = \eta \delta_{ij}
\end{equation}
This reduction uses only the 120-degree rotation matrix and does not invoke a vertical mirror operation. An antisymmetric part of the tensor, which the three-fold axis allows, is orthogonal to the symmetric strain tensor and drops out of Eq.~(3). The reduction to a single scalar $\eta$ therefore survives the lowering of the idealized $C_{3v}$ symmetry to the $C_3$ symmetry of the relaxed Hollow and Chromium-aligned registries. Because $M$ is derived from magnetic exchange, Neumann's principle should in general be applied with the magnetic point group and not the structural point group. Here the distinction does not matter. The substrate magnetization is taken along the out-of-plane axis, the known easy axis of CrBr$_3$ \cite{tsubokawa1960magnetic, zhang2019direct}, and every operation of the confirmed $C_3$ point group is a rotation about that axis. Each operation therefore preserves the magnetic moment without being combined with time reversal, so the magnetic and structural point groups coincide for this heterostructure and the scalar reduction is the same under either. The Bromine-aligned structure, with space group P1, retains independent in-plane components, and no scalar coupling constant is extracted for it (Sec. III.F).

The density functional theory calculations with the D3 correction bound the coupling at small tensile strain to $|\eta| \lesssim 2$ meV, and the baseline fit gives $\eta = +0.56$ meV. The fit uses a strain series from zero to $+1$\% applied strain in steps of 0.25\%, with the ions relaxed at each strain to a force threshold of 5 meV/\AA\ before the spin-orbit calculation. The splitting $M$ at each strain is $M = (\Delta E_{C1} - \Delta E_{C2})/2$, where $\Delta E_{C1}$ and $\Delta E_{C2}$ are the $K$--$K'$ energy differences of the lower and upper spin partner of the $\text{MoS}_2$ conduction-band doublet. The two partners have opposite spin, so their energy differences have opposite signs, and $M$ is the average of their magnitudes with the sign of $\Delta E_{C1}$. Regression through these five points gives a slope of $+0.0112$ meV per percent strain with a standard error of 0.0085 meV per percent. The actuator strain is tensile and small, so the slope on the tensile side of zero strain is the relevant one. The DFT strain sweep is equal-biaxial, so $\varepsilon_{xx}=\varepsilon_{yy}$ and $M = M_0 + \eta(\varepsilon_{xx}+\varepsilon_{yy})$ gives $dM/d\varepsilon_{applied} = 2\eta$. Converting the slope to fractional strain ($\times 100$) and dividing by two gives a nominal $\eta = +0.56$ meV with a 95\% confidence interval of $[-0.79, +1.91]$ meV, which brackets zero because the fitted shift of about 11 $\mu$eV over the 1\% sweep lies at the numerical precision of the calculation.

Across the cutoff, basis-set, and mesh settings of Sec. IV.A, the two-point estimates of $\eta$ range from $-0.7$ to $+0.9$ meV, reflecting a numerical precision of about 15 $\mu$eV in the splitting differences. Since the actuator strain of $3.5 \times 10^{-4}$ shifts the splitting by only 0.2 $\mu$eV, nearly two orders of magnitude below this precision floor, the first-principles calculations can bound $|\eta|$ to about 2 meV but cannot resolve its value or sign. The transport analysis below addresses this gap. At $T = 10$ K the 3.2 meV conduction-band spin-orbit splitting confines about 98\% of the thermal occupation to the lower spin partner. Fitting $\Delta E_{C1}$ alone over the same 0 to $+1$\% window (where its computed 0\% value is $-0.469$ meV and the fitted intercept is $-0.473$ meV) gives a slope of $+0.0188 \pm 0.0067$ meV per percent strain, which corresponds to $\eta = +0.94$ meV and lies within the same $|\eta| \lesssim 2$ meV bound. The doublet-averaged baseline value $|\eta| = 0.56$ meV is retained below as a conservative example target, since the resolution threshold scales as $1/|\eta|$.

The slope depends on the strain window, because $M$ is not linear in strain over a few percent (Sec. IV.B). Fits over other windows give $\eta$ between $-5.4$ and $+0.6$ meV (Table S5 of the Supplemental Material \cite{SM}). The window from 0 to $+1$\% is the tensile window that matches the actuator, for the reason given above. The series also includes compressive points and tensile points above $+1$\%, which are treated in Sec. IV.B and are not part of this fit.

\subsection{Continuum Piezoelectric Actuation}

The microscopic scalar $\eta$ is passed to a continuum elastodynamic model. The linear piezoelectric constitutive equations \cite{nye1985physical} are given below, with material coefficients ($C_{ijkl}$, $e_{kij}$, $\kappa_{ij}$) consistent with the measured elastic and piezoelectric constants of wurtzite AlN reported by Tsubouchi and Mikoshiba \cite{alnConstants}, as implemented in the COMSOL Material Library. The numerical values are given in the Supplemental Material \cite{SM}.
\begin{equation}
\sigma_{ij} = C_{ijkl} \varepsilon_{kl} - e_{kij} E_k
\end{equation}
\begin{equation}
D_i = e_{ikl} \varepsilon_{kl} + \kappa_{ij} E_j
\end{equation}
\begin{equation}
\frac{\partial \sigma_{ij}}{\partial x_j} = 0
\end{equation}
\begin{equation}
\frac{\partial D_i}{\partial x_i} = 0
\end{equation}
The simulated device is a plane-strain cross-section of an aluminum nitride membrane, 200 $\mu$m long and 1 $\mu$m thick, and the strain trace is evaluated at the top surface, $y = 0.5~\mu$m, where the heterostructure sits. A tensile bias is applied through prescribed boundary displacements. No initial stress is imposed, because with clamped ends it would distort the stress field in the interior of the membrane. The device has a two-terminal split-electrode design and is driven through a nine-state voltage sweep. Within the central active region $x \in [-40, 40]$ $\mu$m, the strain field is spatially uniform for each voltage state, to within 0.01\% of the mean trace. Because the Neumann's-principle derivation reduces $\eta$ to a single scalar, spatial uniformity is sufficient. Identifiability then requires several distinguishable strain magnitudes, not complex asymmetric spatial strain shapes.

\subsection{Valley-Resolved Transport and the Transverse Voltage Observable}

The mechanical response is connected to the transport observable by a two-species drift-diffusion-Poisson model \cite{markowich1986stationary, gajewski1985existence}. The valley-resolved carrier fluxes are
\begin{equation}
J_\tau = e n_\tau \mu_\tau (-\nabla \phi) + e D_\tau \nabla n_\tau + \sigma_\tau^{(AH)} (-\nabla \phi \times \hat{z})
\end{equation}
where $\sigma_\tau^{(AH)} = \tau \beta n_\tau$ is the valley-contrasting anomalous Hall conductivity that produces the transverse open-circuit voltage in Eq.~(13). The steady-state continuity equation includes relaxation toward local equilibrium
\begin{equation}
\nabla \cdot J_\tau + e \left[ G_\tau - R_\tau - \Gamma_{\tau \to \tau'} n_\tau + \Gamma_{\tau' \to \tau} n_{\tau'} \right] = 0
\end{equation}
where the intervalley rates depend on the valley splitting as
\begin{equation}
\begin{split}
\Gamma_{K \to K'} &= \frac{1}{\tau_v} \exp \left( -\frac{\Delta E_{valley}(\varepsilon, \eta)}{2 k_B T} \right), \\
\Gamma_{K' \to K} &= \frac{1}{\tau_v} \exp \left( \frac{\Delta E_{valley}(\varepsilon, \eta)}{2 k_B T} \right).
\end{split}
\end{equation}
At equilibrium the two rates balance, which gives the density ratio $n_K / n_{K'} = r(\varepsilon, \eta) = \exp\left(\Delta E_{valley} / k_B T\right)$. A positive $\Delta E_{valley}$ therefore favors the $K$ valley. With $M(\varepsilon) = E_K - E_{K'}$ as defined in Sec. II.D, the splitting that enters the rates is $\Delta E_{valley} = -M(\varepsilon)$, so that the lower-energy valley is favored.

The electrostatic potential obeys the Poisson equation
\begin{equation}
\nabla \cdot (\kappa \nabla \phi) + e (p - n_K - n_{K'} + N_D^+ - N_A^-) = 0
\end{equation}
As the Hall angle stays below 1\% up to the 10 percent velocity-saturation ceiling of Sec. IV.C and the channel is much longer than the valley diffusion length, the transport problem simplifies under a uniform longitudinal field. The Poisson equation fixes the prescribed field, the anomalous Hall flux acts as a small transverse correction that does not enter the continuity equations, and generation and recombination vanish. The system then reduces to two coupled one-dimensional linear drift-diffusion equations for $n_K$ and $n_{K'}$. The intervalley scattering time is set to $\tau_v = 1$ ns, an order-of-magnitude value from the literature. Section V.B shows that the resolution threshold is unchanged across two decades of $\tau_v$. An open-circuit Hall condition then gives the transverse voltage observable \cite{mak2014valley}
\begin{equation}
V_H(x) = -W E_{field} \frac{\beta}{e \mu} \frac{n_K - n_{K'}}{n_K + n_{K'}}
\end{equation}
The scalar parameter $\beta$ is the leading-order ($k=0$) evaluation of the Berry curvature, an analytical approximation to the complete Brillouin-zone thermal integral. It is computed as
\begin{equation}
\beta = \frac{e^2}{\hbar} 2 a^2 \left( \frac{t}{\Delta} \right)^2
\end{equation}
The measurement model includes Johnson-Nyquist thermal noise and an amplifier noise floor of $v_n = 5$ $\text{nV}/\sqrt{\text{Hz}}$, at a fixed temperature of $T = 10$ K. This lies below the Curie temperature of $\text{CrBr}_3$, roughly 30 to 40 K \cite{tsubokawa1960magnetic, zhang2019direct}, so the substrate is magnetically ordered and the proximity splitting can exist. The zero-strain splitting and the coupling are density functional theory values with the moment fixed at 6.00 $\mu_B$, which corresponds to the fully polarized substrate, and the thermal reduction of the substrate magnetization multiplies both (Sec. V.D). The conversion from three-dimensional to two-dimensional electronic density uses an effective electrical thickness of $d_{thickness} = 0.65$ nm \cite{radisavljevic2011single}. This thickness is a model input. The $N_s$ Hall probes tile the channel, each an electrode pair of length $\ell_p$ equal to the sensor pitch, and the Johnson noise of a probe is that of the resistance $R = W/(\ell_p \sigma_{xx})$ between its electrodes across the channel width, with $\sigma_{xx}$ the sheet conductance.

\section{The Inverse Problem and Identifiability Framework}

The forward model is inverted to map macroscopic device transport back to the microscopic strain coupling. This section sets up the inversion and gives the limits on how well $\eta$ can be identified.

\subsection{The Forward Operator and its Differentiability}

The inversion needs a forward operator that maps the coupling parameter $\eta$ to a predicted device observable. Let $u = (n_K, n_{K'}, \phi)$ be the state variables of the drift-diffusion-Poisson system. In the regime treated below the potential reduces to the prescribed uniform field, so the state is $(n_K, n_{K'})$. An observation operator $\mathcal{O}$ extracts the transverse Hall potential difference $V_H$ at discrete sensor locations $x_m$, which gives the parameter-to-observable map $\mathcal{P}[\eta] = \mathcal{O}[u(\eta)]$. The implicit function theorem justifies treating this map as continuously Fr\'echet differentiable in a neighborhood of the operating solution \cite{lions1971optimal}. This ensures that the gradient used by the inversion exists.

\subsection{The Adjoint-State Method}

Finite differencing needs one extra forward solve for each parameter dimension. The adjoint-state method gives the complete gradient with a single additional linear partial differential equation solve, regardless of the number of parameters \cite{tarantola2005inverse}. 

The forward constraint is the abstract linear system $A(\eta)u = b$, which represents the steady-state transport physics. The Lagrangian is $\mathcal{L} = J(\eta) + \langle \lambda, A(\eta)u - b \rangle$, where $J$ is the objective functional and $\lambda$ is the adjoint variable. Stationarity with respect to the state variable gives $A^T \lambda = -\nabla_u J$. Stationarity with respect to the parameter then gives the abstract gradient $dJ/d\eta = \lambda^T [\partial A/\partial \eta] u + \alpha \eta$. With the positive forcing convention $A^T \lambda = \nabla_u J$ for the continuous adjoint equation, the gradient becomes $dJ/d\eta = \alpha \eta - \lambda^T [\partial A/\partial \eta] u$. This shows where the negative sign comes from and fixes the convention used in all computational routines.

The gradient is verified against a finite-difference reference, and its relative error falls by a factor of 100 for each factor of 10 in step, from $3.4 \times 10^{-3}$ at $10^{-2}$ to $3.4 \times 10^{-9}$ at $10^{-5}$, as expected for a central difference, and reaches $1.8 \times 10^{-10}$ at $10^{-6}$, where rounding limits it. The numerical values, together with the discretization and solver details, are in the Supplemental Material \cite{SM}.

\subsection{Fisher Information and the Cram\'er-Rao Bound}

Statistical identifiability needs a measurement noise model. For independent Gaussian measurement noise with covariance $\Sigma$, the log-likelihood gives the Fisher information matrix $I(\eta) = K(\eta)^T \Sigma^{-1} K(\eta)$, where $K$ is the matrix of sensitivities of the sensor readings to the parameters (a single column here, since $\eta$ is the only parameter). The Cram\'er-Rao bound is the minimum achievable covariance of any unbiased estimator of the coupling parameter \cite{cramer1946mathematical, rao1945information}. In other words, it sets a floor on how small the uncertainty of any unbiased estimate of $\eta$ can be for a given measurement. It is an asymptotic, linearized bound, and it also governs parameter identifiability. The parameter $\eta$ is locally identifiable if and only if the Fisher information matrix is nonsingular, which is equivalent to $K(\eta)$ having full column rank.

\subsection{Regularized Nonlinear Least Squares}

The inversion minimizes a nonlinear least-squares objective, with the Tikhonov term $\alpha\eta^2/2$ of Sec. III.B available for regularization. Since $\eta$ is a low-dimensional scalar and not a distributed spatial field, the problem is well conditioned, the regularization weight is set to zero in the computations below, and the existence and uniqueness arguments are simpler than for infinite-dimensional inverse problems. Minimization uses Levenberg-Marquardt updates, which have established local convergence guarantees provided the sensitivity matrix has full rank and the residual misfit is sufficiently small \cite{nocedal2006numerical}. A recovery test on synthetic data (Sec. S10 of the Supplemental Material \cite{SM}) shows that the estimator reaches the Cram\'er-Rao bound within the sampling uncertainty.

\subsection{D-Optimal Experimental Design}

The Fisher information matrix depends on the spatially inhomogeneous strain field produced by the piezoelectric actuator, so the electrode layout is an experimental design variable. D-optimal design maximizes the determinant of the Fisher information matrix, which minimizes the volume of the asymptotic confidence ellipsoid of the recovered parameters \cite{kiefer1959optimum}. For the single scalar $\eta$ the determinant is the Fisher information itself, so the criterion selects the actuation state with the largest sensitivity. The criterion is applied to the nine discrete actuation states of the device, which are compared in pairs through the determinant of the Fisher information matrix for $M_0$ and $\eta$. The quantitative results are given in Section IV.

\subsection{Symmetry-Collapse Illustration}

A short analytical illustration shows how identifiability depends on the point-group symmetry. The Bromine-aligned registry has space group P1 and therefore retains independent in-plane coupling components, which the illustration represents by the independent constants $\eta_{xx}$ and $\eta_{yy}$ driven by an actuation pattern with a fixed linear strain ratio. The two-by-two Fisher information matrix of this system is exactly rank-deficient for any single actuation state of fixed strain ratio. The determinant is identically zero, a closed-form rank deficiency and not a numerical near-singularity. In the computed illustration the strain ratio is fixed at $\varepsilon_{yy} = -0.27\,\varepsilon_{xx}$ (taken from Bertolazzi et al. \cite{bertolazzi2011stretching}), and the determinant evaluates to $-9.5 \times 10^{5}$ against $F_{11}F_{22} \approx 4.6 \times 10^{21}$, which is zero to floating-point precision. Intuitively, with $\varepsilon_{yy}$ locked to $-0.27\,\varepsilon_{xx}$, the data depend only on the combination $\eta_{xx} - 0.27\,\eta_{yy}$, so $\eta_{xx}$ and $\eta_{yy}$ cannot be separated. The confirmed $C_3$ symmetry of the Hollow and Chromium-aligned registries sets $\eta_{xx} = \eta_{yy} = \eta$ and removes this rank deficiency. The P3 symmetry of the Hollow registry, which is the lowest-energy registry with the D3 correction, therefore underpins the single-scalar identifiability analysis. The symmetry of each registry is determined by the structural relaxation in Section II, and the collapse is plotted in Section IV.

\section{Results}

The framework is applied to the molybdenum disulfide and chromium tribromide heterostructure to evaluate the viability of strain-tuned valley splitting. The pipeline passes the atomic parameters through continuum elastodynamics to obtain bounds on device-level identifiability.

\subsection{Density Functional Theory Parameterization and Validation}

\begin{figure}[htbp]
    \centering
    \includegraphics[width=\columnwidth]{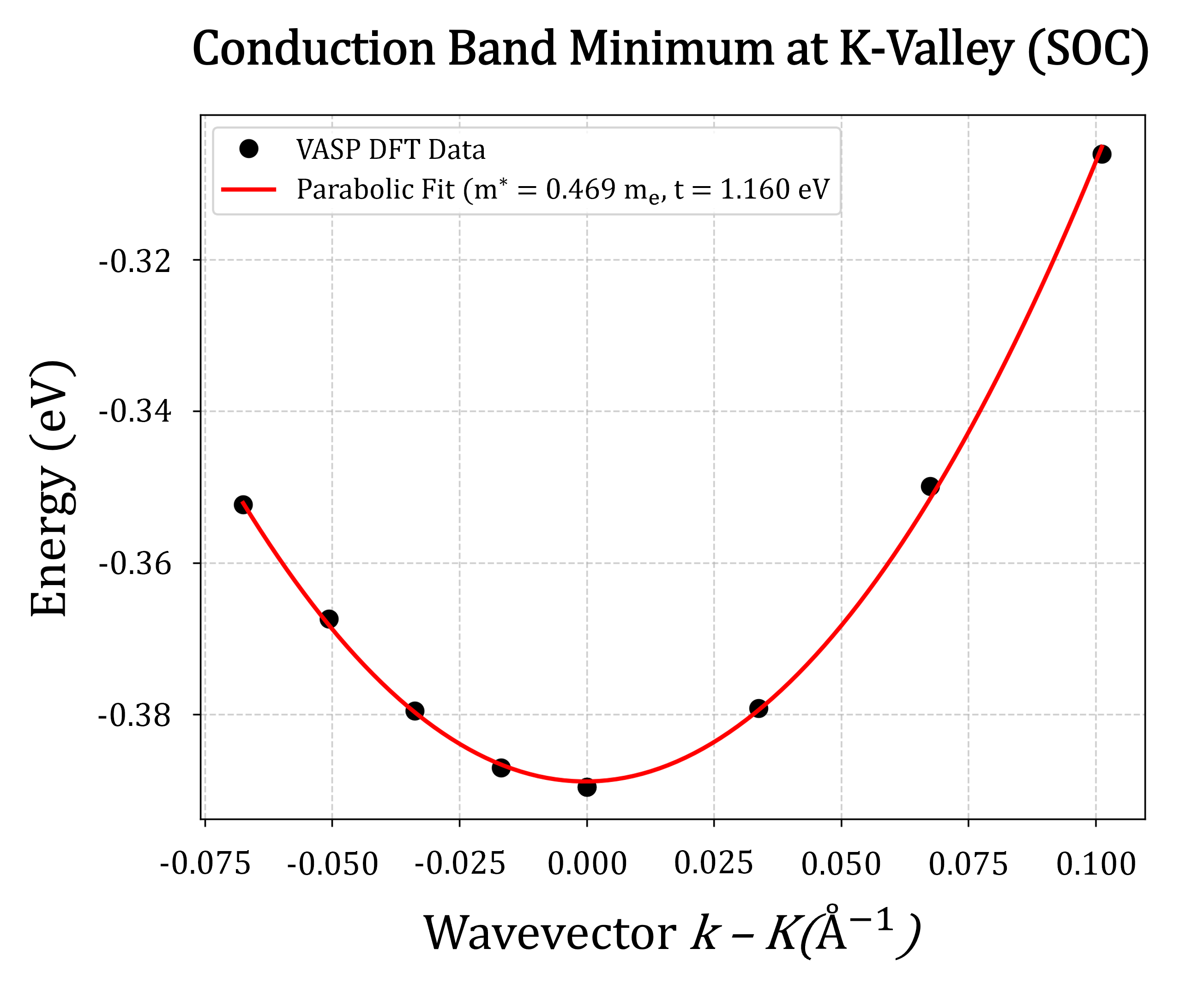}
    \caption{Parabolic fit to the $K$-valley conduction band minimum of the transition metal dichalcogenide monolayer. The regression gives $m^* = 0.4686 m_e$ and $t = 1.1601$ eV over the narrowed 9-point fit window.}
    \label{fig:band_curvature}
\end{figure}

\begin{figure}[htbp]
    \centering
    \includegraphics[width=\columnwidth]{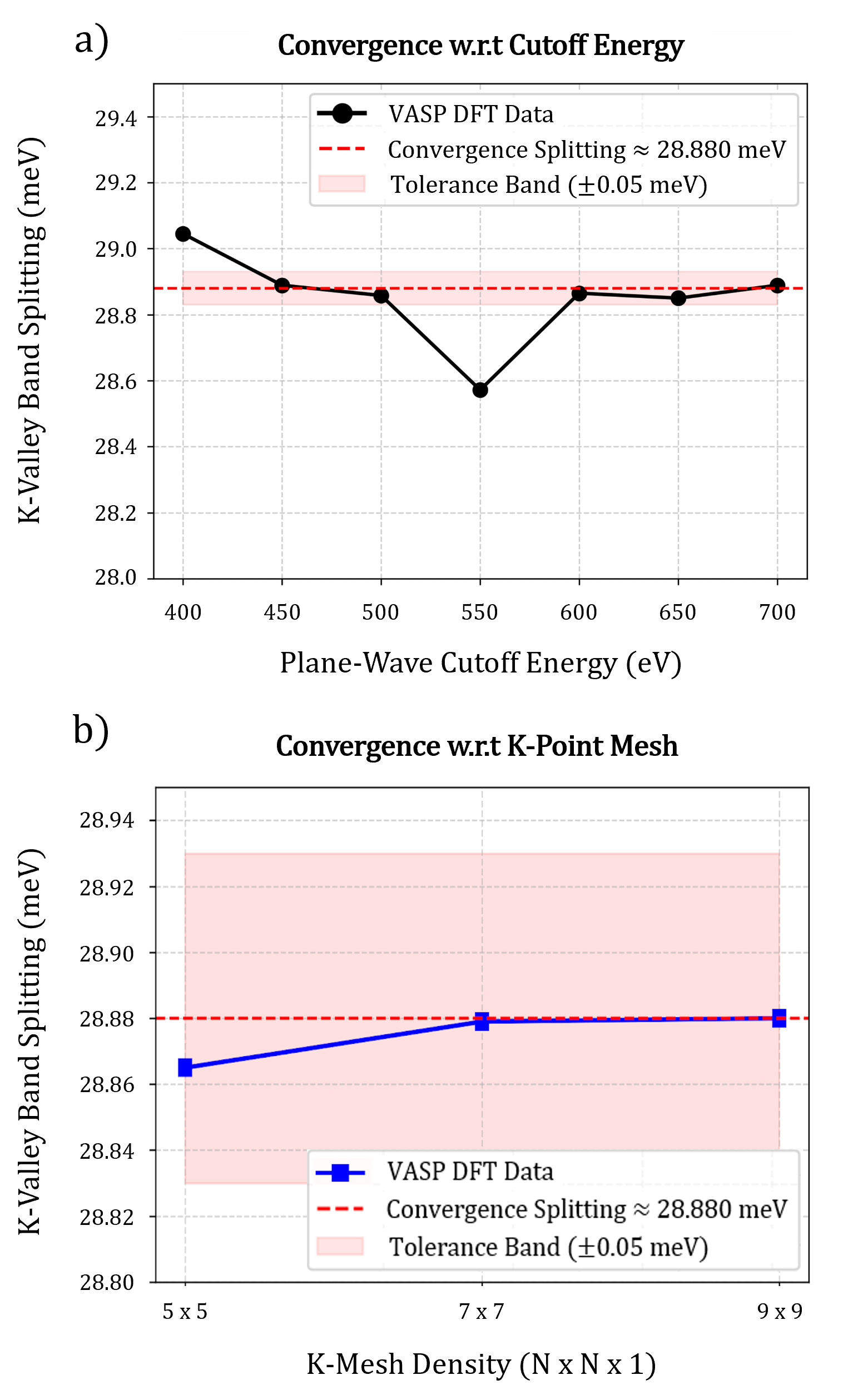}
    \caption{Convergence of a representative substrate-hybridized eigenvalue splitting with (a) plane-wave cutoff and (b) $k$-point mesh (the unconverged $3 \times 3$ point at 3.746 meV is omitted for scale). The shaded band is $\pm 0.05$ meV around the converged value of 28.880 meV.}
    \label{fig:dft_convergence}
\end{figure}

A parabolic fit restricted to the $K$-valley conduction band minimum gives an effective mass of $m^* = 0.4686 m_e$ and an effective hopping parameter of $t = 1.1601$ eV (Fig.~\ref{fig:band_curvature}). The fit window was narrowed from 11 to 9 points to isolate the parabolic pocket and avoid the onset of trigonal warping at the Brillouin zone boundary. An energy convergence sweep on a representative substrate-hybridized eigenvalue tests numerical stability. For plane-wave cutoffs from 600 to 700 eV, the splitting stays within the $\pm 0.05$ meV tolerance band around the converged value of 28.880 meV, and for $\Gamma$-centered meshes of $7 \times 7$ and denser it agrees with that value to within 0.001 meV. The 550 eV point lies 0.31 meV below the converged value. The $3 \times 3$ mesh has an unconverged magnetic moment and is omitted from Fig.~\ref{fig:dft_convergence}(b). The production calculations use a cutoff of 600 eV.

The splitting itself is tested on the D3 structure at fixed geometry. At 0\% strain $M_0$ is $-0.4256$, $-0.4521$, $-0.4553$, and $-0.4448$ meV at cutoffs of 500, 600, 700, and 800 eV on the $6 \times 6 \times 1$ mesh, and $-0.4543$ meV on a $9 \times 9 \times 1$ mesh at 600 eV. From 600 eV upward it is fixed to 0.01 meV, about 2\% of its value. The difference between $M$ at $+1$\% and at 0\% is a smaller quantity and depends on these choices more strongly. It is $+0.010$ meV at 600 eV, $-0.004$ meV at 700 eV, and $+0.017$ meV on the finer mesh. Keeping the plane-wave basis the same size at each strain, by scaling the cutoff as $(1+\varepsilon)^{-4/3}$, gives $+0.005$ meV from a base cutoff of 600 eV and $-0.013$ meV from 800 eV. The corresponding two-point estimates of $\eta$ lie between $-0.7$ and $+0.9$ meV. Since the 1\% strain sweep shifts the splitting by only about 10 $\mu$eV, comparable to the $\pm 15~\mu$eV numerical precision set by these choices, and the actuator shifts it by only 0.2 $\mu$eV, first-principles theory fixes $M_0$ to about 2\% and bounds the magnitude of $\eta$ while leaving its value and sign to a transport measurement. Section IV.C determines the conditions under which that measurement can resolve them. The full sweep of the eigenvalue is in Fig.~\ref{fig:dft_convergence}.

\subsection{Extraction of the Exchange-Strain Coupling}

The coupling constant is extracted from strain series computed with the D3 correction, using the signed splitting $M$ at each strain (Table S3 of the Supplemental Material \cite{SM}). 

\begin{figure}[htbp]
    \centering
    \includegraphics[width=1\columnwidth]{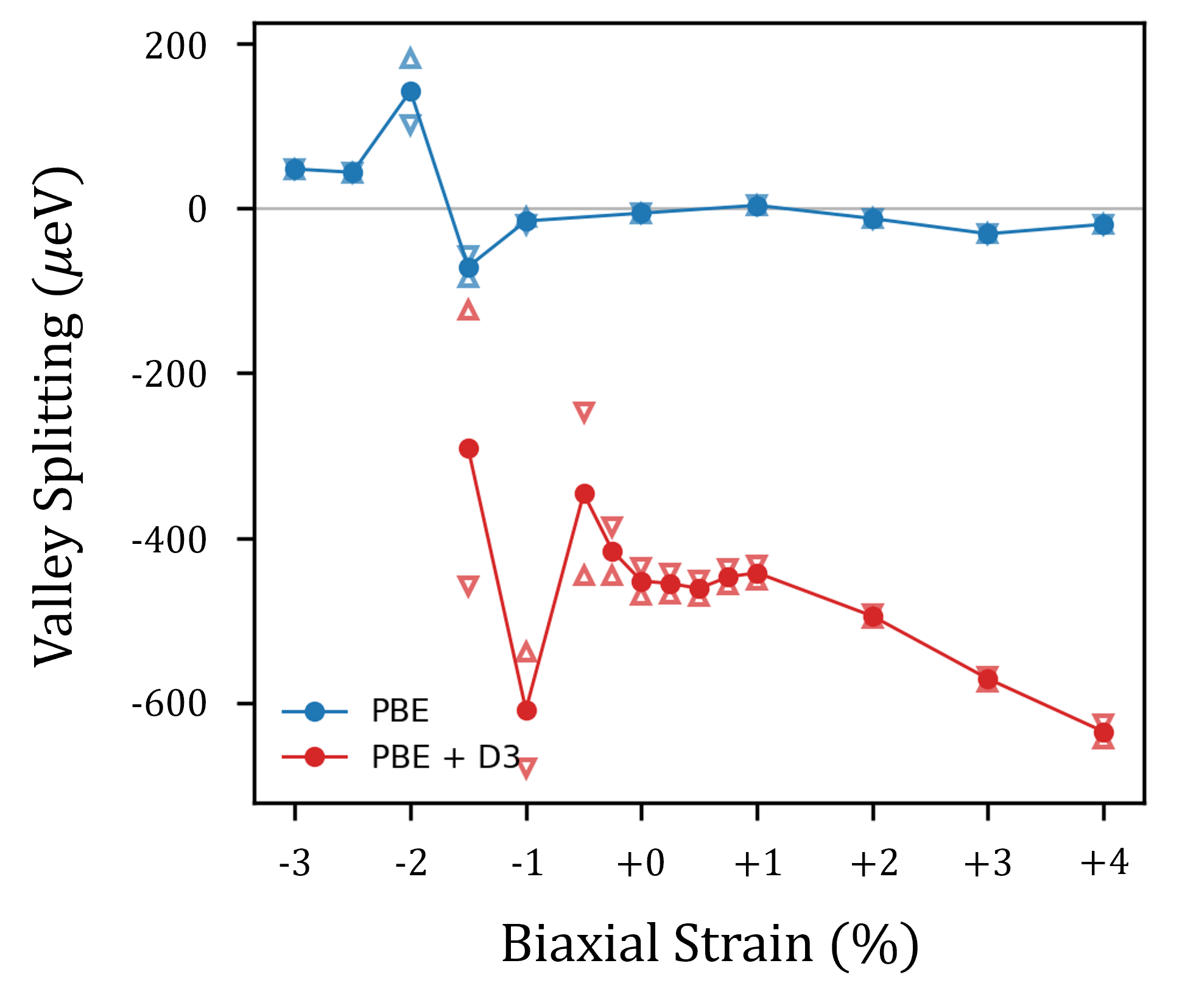}
    \caption{Signed valley splitting $M$ of the $\text{MoS}_2$ conduction-band doublet against applied biaxial strain, with the DFT-D3 correction (red) and without it (blue). Open triangles are the $K$--$K'$ energy differences of the two spin partners ($\Delta E_{C1}$ and $-\Delta E_{C2}$), and circles are their average $M$. The D3 point at $-2$\% is not shown because the doublet is not identified there.}
    \label{fig:M_vs_strain}
\end{figure}

From 0 to $+1$\% in steps of 0.25\%, $M$ varies between $-0.442$ and $-0.461$ meV. A least-squares line through these five points has a slope of $+0.0112 \pm 0.0085$ meV per percent strain. Converting to fractional strain and accounting for the factor of two from the equal-biaxial strain protocol ($dM/d\varepsilon_{applied} = 2\eta$, see Section II.D) gives a nominal $\eta = +0.56$ meV with a 95 percent confidence interval of $[-0.79, +1.91]$ meV. The interval brackets zero because the fitted shift of about 11 $\mu$eV over the sweep lies at the numerical precision of the calculation (Sec. IV.A). Together with the convergence tests, the DFT result therefore bounds $|\eta|$ to about 2 meV at tensile strains up to $+1$\%, and the value and sign of the coupling are left to the transport measurement. All subsequent identifiability calculations use $|\eta| = 0.56$ meV, the baseline result, as the target value.

Over a wider range the response is not linear (Fig.~\ref{fig:M_vs_strain}). With D3, $M$ is $-0.608$ meV at $-1$\%, and beyond $+1$\% it falls by about 0.064 meV per percent strain to $-0.6345$ meV at $+4$\%. Lines through 0 to $+4$\% and through $\pm 0.5$\% give $\eta = -2.3$ and $-5.4$ meV (Table S5 of the Supplemental Material \cite{SM}).

On the compressive side the doublet crosses two other bands. The first has $\text{CrBr}_3$ character (weight 0.79, with $\text{MoS}_2$ weight 0.04) and crosses at $-0.64$\% strain, and the second has mixed $\text{MoS}_2$ and $\text{CrBr}_3$ character (weights 0.26 and 0.51) and crosses at about $-1.4$\%. The magnitudes of the two partner splittings differ by 0.20 and 0.33 meV at $-0.5$ and $-1.5$\%, against 18 to 34 $\mu$eV from 0 to $+1$\%, and the plateau from 0 to $+1$\% lies at least 60 meV from any band with $\text{CrBr}_3$ weight above 0.1. At $-2$\% the doublet at $K'$ is spread over three bands and is not identified. Without the D3 correction the splitting is smaller by a factor of 75 at zero strain, and the mixed band crosses the doublet near $-1.8$\% strain, where $M$ changes sign from $-0.071$ to $+0.142$ meV.

\subsection{The Device-Level Identifiability Limit}

This section determines the field conditions under which a coupling of a given magnitude is statistically resolvable. A coupling counts as resolved when the 95\% confidence interval implied by the Cram\'er-Rao bound excludes zero, that is, when $1.96\,\sigma_\eta < |\eta|$, and the resolution threshold is the lowest field at which this holds. Taking $|\eta| = 0.56$ meV from Section II.D as an example, the analysis determines the longitudinal electric field required for a coupling of this magnitude to be resolved above the measurement noise floor. The calculation uses the continuous adjoint sensitivity and a field-dependent mobility model. The mobility follows the Caughey-Thomas high-field saturation model \cite{caughey1967carrier} with $v_{sat} = 2.0 \times 10^6$ cm/s, which is below the room-temperature value of $(3.4 \pm 0.4) \times 10^6$ cm/s inferred, after correcting for self-heating, for synthetic monolayer MoS$_2$ \cite{smithe2018high}. The lower $v_{sat}$ places the trust boundaries at lower fields than the measured value would. The low-field mobility $\mu_0$ and $v_{sat}$ are held at the values listed in Table~\ref{tab:model_parameters}.

The actuator strain field is not equal-biaxial. The ratio $\varepsilon_{yy}/\varepsilon_{xx}$ ranges from $-0.17$ to $-0.33$ across the nine voltage states and equals $-0.330$ for the state ($V_{\mathrm{inner}} = +10$ V, $V_{\mathrm{outer}} = -10$ V) used for the operating window. The strain trace is therefore taken directly from the finite-element solution and not from an assumed Poisson ratio. 

Against the thermal and amplifier noise floors at 10 K, this transport model gives a statistical resolution threshold of 5.59 V/cm. The threshold is computed for $N_s = 200$ Hall probes of length $\ell_p = 0.4$ $\mu$m, equal to their pitch, and a detection bandwidth of $B = 1$ Hz. The Johnson noise of a probe is about 20 times the amplifier noise, so the threshold scales as $\sqrt{B}$ and is nearly independent of the number of probes that tile the channel (changing by 0.5\% between 100 and 400 probes), because the combined noise of $N_s$ tiled probes matches that of a single electrode pair along the full channel length (Sec. S1 of the Supplemental Material \cite{SM}).

\begin{figure}[htbp]
    \centering
    \includegraphics[width=\columnwidth]{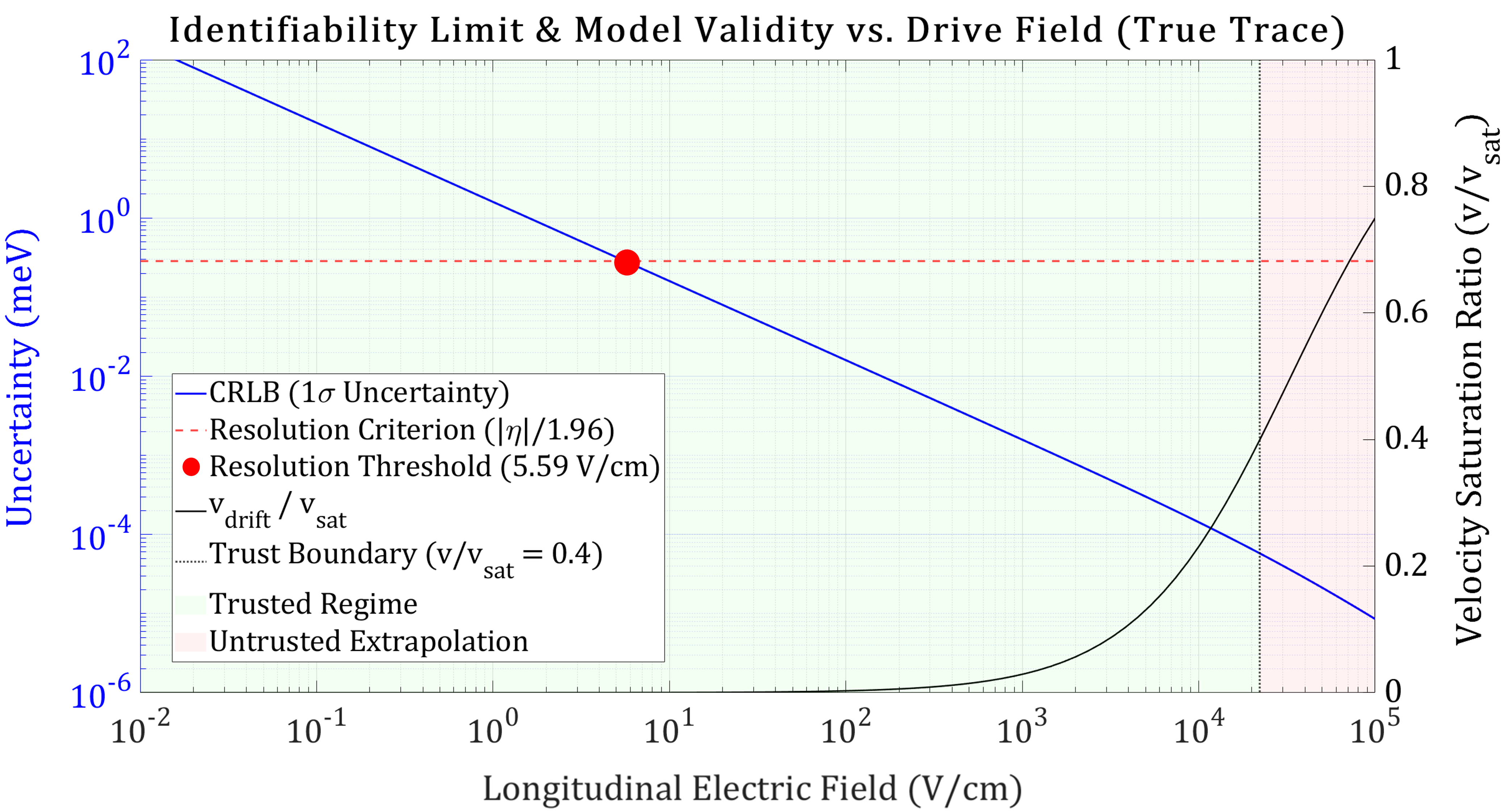}
    \caption{Device-level identifiability and trust boundary limits as a function of the applied longitudinal electric field. The blue curve is the CRLB on $\eta$ ($1\sigma$), the red dashed line is the 95\% resolution criterion $|\eta|/1.96 = 0.286$ meV, and the black dotted line marks the 40 percent trust boundary $v/v_{sat} = 0.4$, an extended high-field reference above the 10 percent ceiling at 3,704 V/cm. The curve crosses the criterion at 5.59 V/cm.}
    \label{fig:trust_boundary}
\end{figure}

The upper bound of the window is set by the velocity-saturation convention. The 10 percent convention, at which the drift velocity reaches 10\% of $v_{sat}$, gives the primary conservative ceiling at 3,704 V/cm, which corresponds to about 30 V across the 80 $\mu$m channel. The 20 and 40 percent conventions common in semiconductor transport modeling \cite{sze2006physics} give extended high-field references at 8,333 V/cm and 22.2 kV/cm, the latter corresponding to about 178 V across the channel. These fields mark the onset of deviation from Ohmic transport. Relative to the 5.59 V/cm floor, the primary ceiling is about 660 times higher, and the 20 and 40 percent references are about 1,500 and 4,000 times higher. The device therefore has an operating window between 5.6 and 3,704 V/cm (Fig.~\ref{fig:trust_boundary}). Below 5.6 V/cm the valley signal cannot be resolved against the noise floor, and above the ceiling the linear transport assumptions weaken progressively. The two bounds arise from independent physical mechanisms. At the 5.6 V/cm lower bound the potential drop across the 80 $\mu$m channel is about 45 mV (or 0.20 V at the 25 V/cm joint-recovery threshold), roughly 50 times $k_BT/e$ at 10 K, while the 3,704 V/cm ceiling corresponds to 30 V (or 0.37 V/$\mu$m in scaled short-channel devices). The resolvable window therefore spans the full practical operating range of two-dimensional transistor channels, from low-bias linear readout to high-field saturation drive.

At the 10 percent ceiling of 3,704 V/cm the Cram\'er-Rao bound is $\sigma_\eta = 413$ neV for the noise floors and probe geometry above, which is about 700 times below the 0.286 meV resolution criterion, and it falls to 58 neV at the 40 percent reference of 22.2 kV/cm. This is the Cram\'er-Rao limit for white Johnson and amplifier noise at the first-principles zero-strain splitting $M_0$, with the Hall gain and the intervalley scattering time known. Section IV.I recovers $\eta$ jointly with all three.

The strain-induced perturbation to the valley splitting satisfies $\eta(\varepsilon_{xx}+\varepsilon_{yy})/k_BT \ll 1$ for every coupling magnitude considered here, staying below $10^{-3}$ for $|\eta| = 2$ meV at 10 K. The sensitivity entering the Fisher information is therefore well approximated by its linear (small-signal) limit about the zero-strain equilibrium. That equilibrium has $|M_0|/k_BT = 0.53$ at 10 K, which corresponds to an equilibrium polarization of 0.26. As a result, the CRLB curve $\sigma_\eta(E)$ in Fig.~\ref{fig:trust_boundary} does not depend on the assumed true value of $\eta$. For any coupling magnitude at or above 0.56 meV, including the upper bound of about 2 meV, the 5.59 V/cm threshold is sufficient, since $\sigma_\eta(5.59\text{ V/cm}) \approx 0.286$ meV does not exceed $|\eta|/1.96$ for any such value. The threshold scales as $1/|\eta|$, so it is about 1.6 V/cm for $|\eta| = 2$ meV and about 16 V/cm for $|\eta| = 0.2$ meV.

\begin{figure}[b]
    \centering
    \includegraphics[width=\columnwidth]{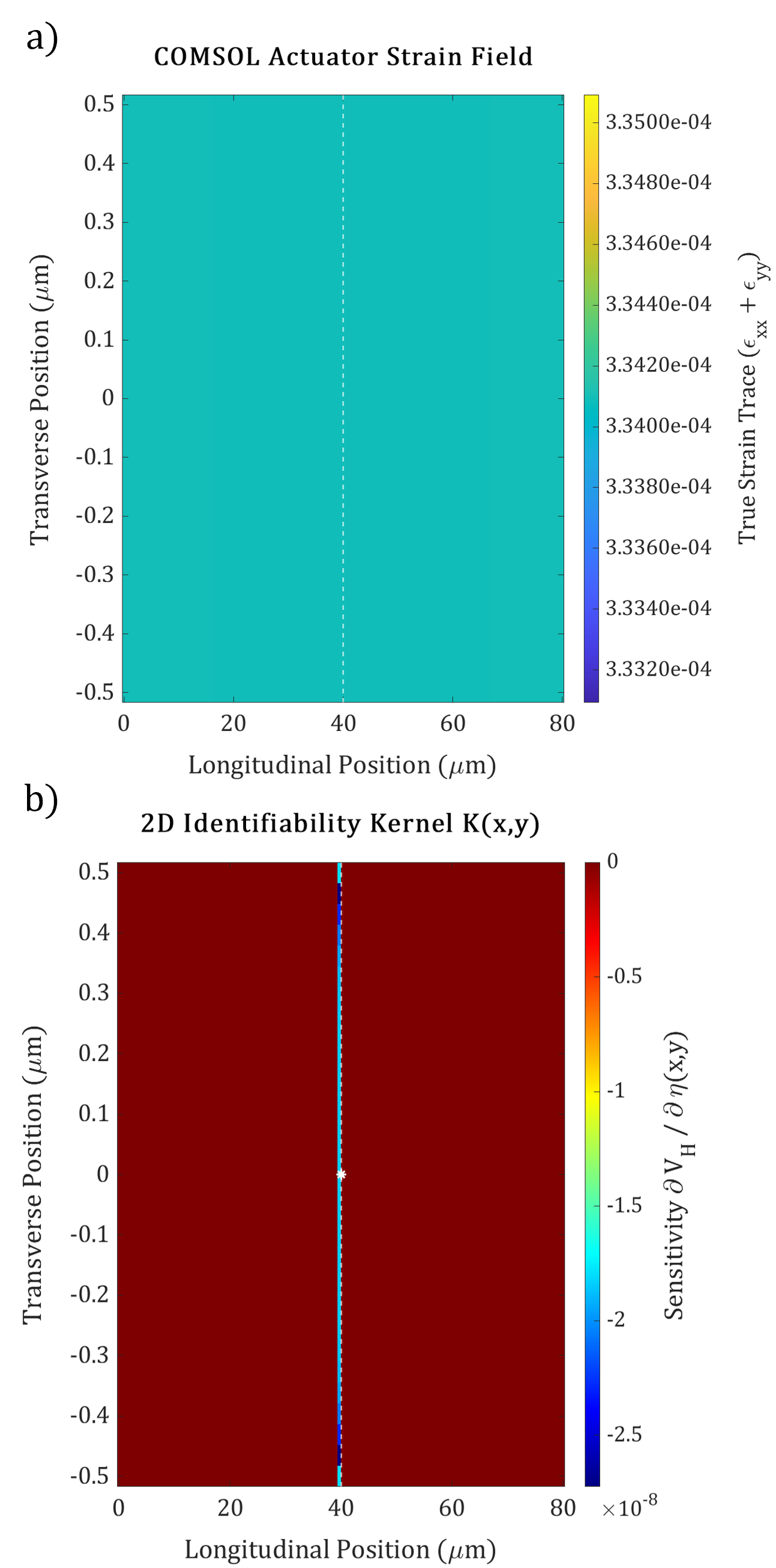}
    \caption{(a) Two-dimensional mechanical strain trace exported from the continuum elastodynamic actuator simulation, showing spatial uniformity across the central device channel. The color scale spans $\pm 1 \times 10^{-6}$ around the mean trace. (b) Two-dimensional sensitivity kernel evaluated with the continuous adjoint method. The sensitivity is localized at the sensing column (dashed line, $x = 40~\mu$m from the contact, the channel midpoint) and spans the channel width.}
    \label{fig:spatial_sensitivity}
\end{figure}

\subsection{Spatial Sensitivity and Actuator Validation}

The continuum piezoelectric simulations show that the strain field across the 80 $\mu$m active region (from $-40$ to $40$ $\mu$m about the channel midpoint) is spatially uniform for every voltage state, to within 0.01\% of the mean trace. The strain trace spans $3.34 \times 10^{-4}$ (both electrodes at $+10$ V) to $4.12 \times 10^{-4}$ (both at $-10$ V) across the nine actuation states in this region and is plotted in Fig.~\ref{fig:spatial_sensitivity}(a). The sensitivity kernel in Fig.~\ref{fig:spatial_sensitivity}(b) is localized at the sensing column at the channel midpoint and spans the channel width, because the valley diffusion length (below $0.23~\mu$m) is far shorter than the channel. The sensitivity pattern is symmetric about the longitudinal axis, so the macroscopic transport observable captures the localized microscopic perturbation. Panels (a) and (b) are computed for the uniform $+10$/$+10$ V state at $E = 1.25$ V/cm and $W = 1~\mu$m.

\subsection{D-Optimal Actuation Design}

\begin{figure}[htbp]
    \centering
    \includegraphics[width=\columnwidth]{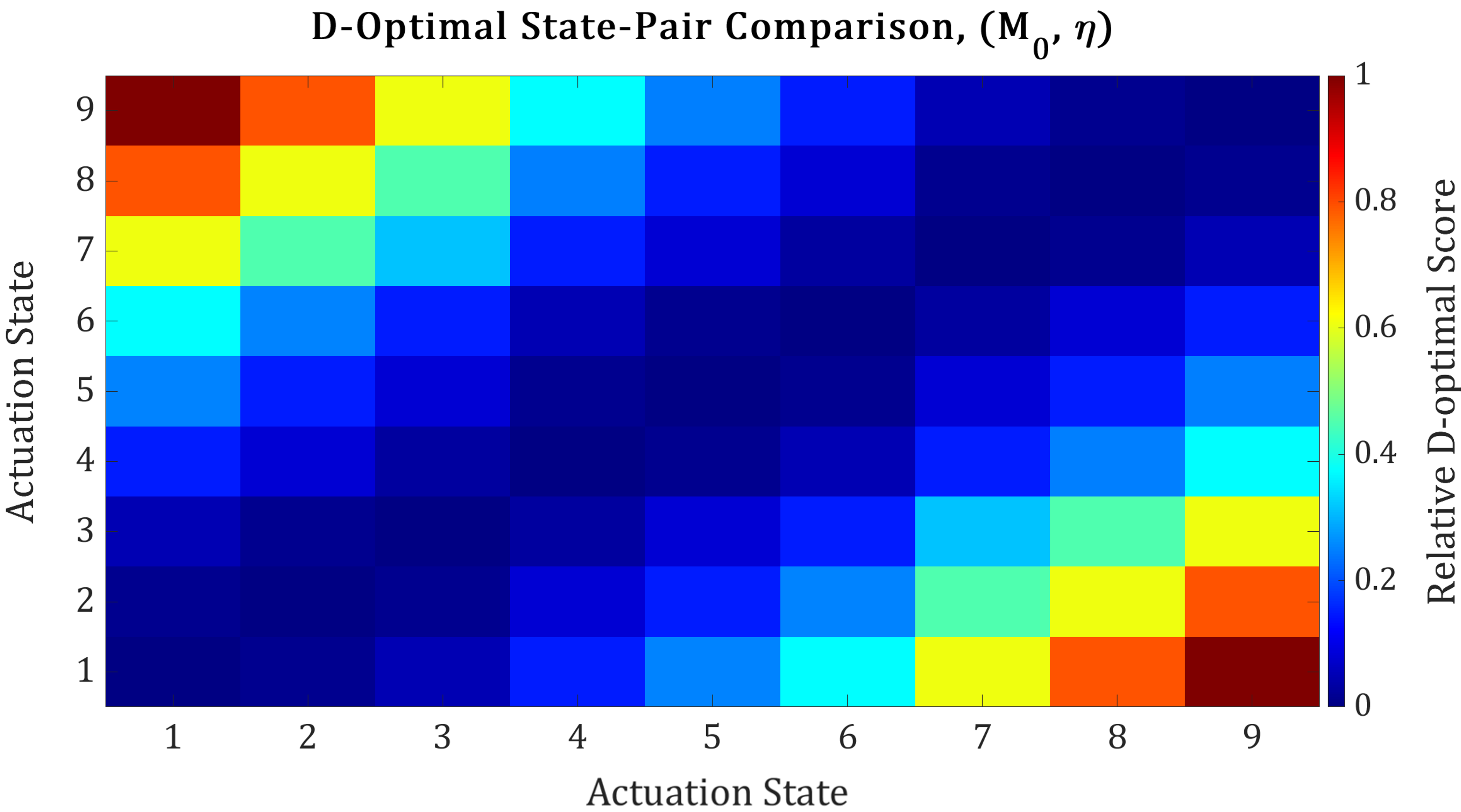}
    \caption{D-optimal comparison of all pairs of actuation states for the joint Fisher information on $(M_0, \eta)$ at 100 V/cm. Each score is normalized to the best pair, states 1 and 9, which has the largest strain contrast.}
    \label{fig:d_optimal}
\end{figure}

The D-optimal comparison scores each pair of actuation states by the determinant of the Fisher information matrix for $(M_0, \eta)$, which sets the volume of the joint confidence ellipsoid of the two parameters. When $M_0$ is unknown, the information on $\eta$ comes from the contrast between the two strain traces and not from their absolute amplitude, so the score of a pair scales as the squared difference of the two traces. The pair of states 1 and 9, driven between $V_{\mathrm{inner}} = V_{\mathrm{outer}} = -10$ V and $+10$ V, gives the largest contrast, $0.78 \times 10^{-4}$, and defines the reference score of 1. The pairs (1, 8) and (2, 9) score 0.79, and the pairs (1, 7) and (2, 8) score 0.61, equal to the squared ratios of their contrasts to that of the best pair (Fig.~\ref{fig:d_optimal}). The inner electrode shifts the trace about 3.6 times more than the outer electrode, so the inner-electrode voltage is the control that sets the contrast. For known $M_0$ the single-state score scales as the squared trace, with a ratio of 1.1055 between states 7 and 9 (Sec. S7 of the Supplemental Material \cite{SM}).

\subsection{Generalized Operating Map}

\begin{figure}[htbp]
    \centering
    \includegraphics[width=\columnwidth]{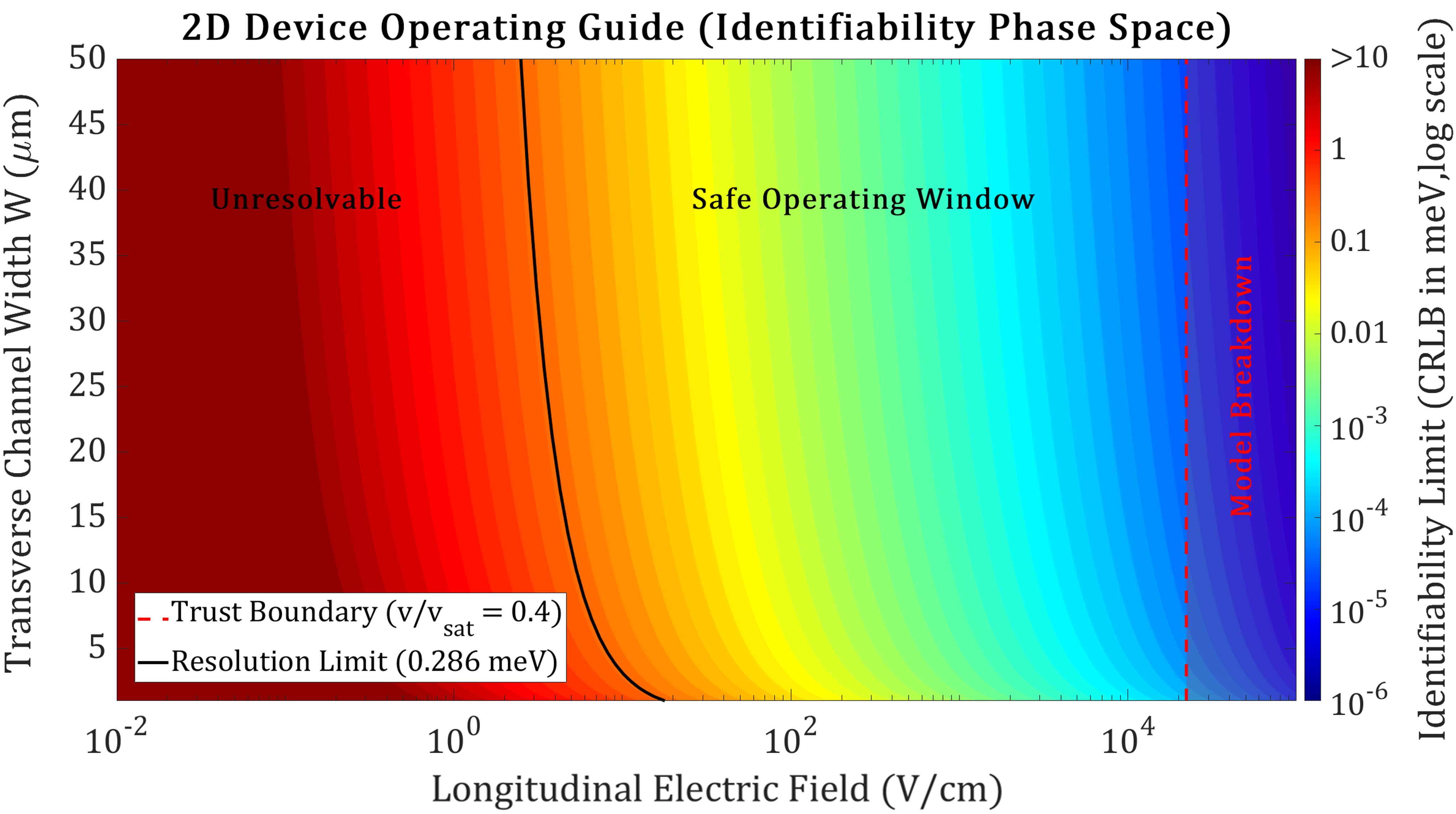}
    \caption{Two-dimensional identifiability map of the statistical resolution limit as a function of the applied longitudinal electric field and the transverse channel width. The black contour is the 0.286 meV resolution limit, and the red dashed line is the $v/v_{sat} = 0.4$ trust boundary, an extended high-field reference above the 10 percent ceiling at 3,704 V/cm.}
    \label{fig:operating_map}
\end{figure}

A two-dimensional operating map shows how the statistical resolution contour scales with the electric field and the channel width together (Fig.~\ref{fig:operating_map}). The contour intersects the 10 $\mu$m channel width slice at approximately 5.6 V/cm, consistent with the one-dimensional result to within the resolution of the sweep grid. The CRLB value at this intersection, $\sigma_\eta \approx 0.286$ meV (equal to $|\eta|/1.96$, the same 95\% resolution criterion used throughout Section IV), is the quantity plotted as the "Resolution Limit" in Fig.~\ref{fig:operating_map}. This agreement is an internal consistency check between the one-dimensional threshold and the two-dimensional pipeline.

\subsection{Symmetry and Identifiability Illustration}

\begin{figure*}[t]
    \centering
    \includegraphics[width=\linewidth]{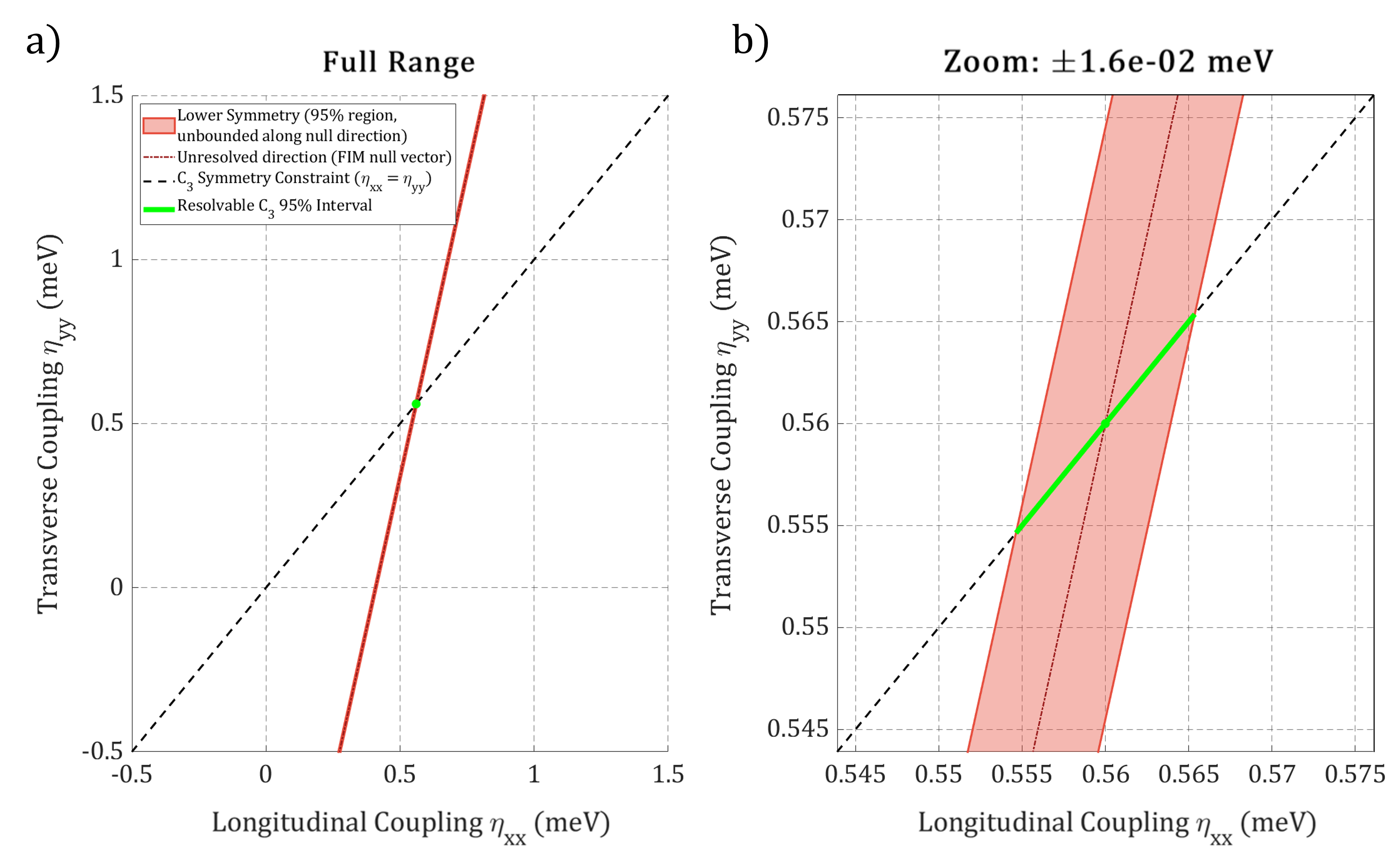}
    \caption{Illustration comparing the bounded confidence interval of the confirmed $C_3$ symmetry class with the singular identifiability collapse of a reduced-symmetry registry such as the Bromine-aligned structure (space group P1). The green segment is the 95\% interval for the single scalar $\eta$, and the red strip is the 95\% region for independent $\eta_{xx}$ and $\eta_{yy}$ under the fixed strain ratio $\varepsilon_{yy} = -0.27\,\varepsilon_{xx}$. The strip is bounded across the identifiable combination $\eta_{xx} - 0.27\,\eta_{yy}$ and unbounded along the null direction of the Fisher information matrix, and no regularization is applied. Panel (a) shows the full range and panel (b) a zoom on the $C_3$ interval. Both are computed at $E = 1000$ V/cm and $W = 10~\mu$m.}
    \label{fig:neumann_proof}
\end{figure*}

Fig.~\ref{fig:neumann_proof} compares the confirmed $C_3$ symmetry of the Hollow registry with the independent in-plane components of the Bromine-aligned registry, which has space group P1 and is represented by the constants $\eta_{xx}$ and $\eta_{yy}$. For a single actuation state of fixed linear strain ratio, the Fisher information matrix of the P1 case is singular, the ratio of whose smallest to largest eigenvalue is of magnitude $1.4 \times 10^{-17}$, and its 95\% confidence region is unbounded along the null direction. The $C_3$ structure has a bounded confidence interval of $\pm 5.4~\mu$eV at this field. This illustrates the analytical result of Section III. For a single actuation state with a fixed strain ratio, the confirmed P3 symmetry removes the unidentifiability that a P1 registry would show.

\subsection{Temperature Dependence of the Operating Window}

The resolution threshold depends on temperature through the thermal energy that sets the equilibrium polarization, through the Johnson noise, and through the magnetization of the substrate. The threshold is evaluated from 2 to 300 K for two substrates. The first stays ordered at every temperature, with $M_0$ and $\eta$ held at their first-principles values. The second is $\text{CrBr}_3$, for which $M_0$ and $\eta$ scale with the reduced magnetization $m(T) = (1 - T/T_c)^{0.3}$ and $T_c = 34$ K \cite{zhang2019direct}. The threshold is evaluated at 1 V/cm, in the Ohmic regime where the Cram\'er-Rao bound scales as $1/E$. The mobility, the saturation velocity, and the intervalley scattering time are held at the values of Table~\ref{tab:model_parameters} at every temperature, and the trust boundaries are those of Sec. IV.C.

For the ordered substrate the threshold stays between 1.7 and 2.4 V/cm below 5 K, equals 5.59 V/cm at 10 K, and rises to 860 V/cm at 300 K, a factor of 154 above its 10 K value (Fig.~\ref{fig:temperature}). The full-field calculation of Table S8 of the Supplemental Material \cite{SM}, which includes the field dependence of the mobility at the threshold, gives 850 V/cm at 300 K. The window at 300 K therefore spans 860 to 3,704 V/cm, a factor of 4.3 between the two bounds, and a coupling of 0.56 meV is resolvable at room temperature in a substrate that stays ordered there. When $\eta$ is recovered jointly with the zero-strain splitting, the Hall gain, and the intervalley scattering time, the threshold at 300 K rises to $3.72 \times 10^{3}$ V/cm (Table S8), essentially at the 10 percent ceiling, so that the joint window at 300 K opens just above the ceiling and extends to the 20 and 40 percent references at 8,333 V/cm and 22.2 kV/cm. For $\text{CrBr}_3$ the threshold follows the ordered case at low temperature and equals 6.1 V/cm at 10 K and 19.8 V/cm at 20.2 K. It reaches 146 V/cm at 33.78 K, 25 times below the 10 percent ceiling, and diverges at $T_c$, where the substrate magnetization vanishes and the window closes.

\begin{figure}[htbp]
    \centering
    \includegraphics[width=\columnwidth]{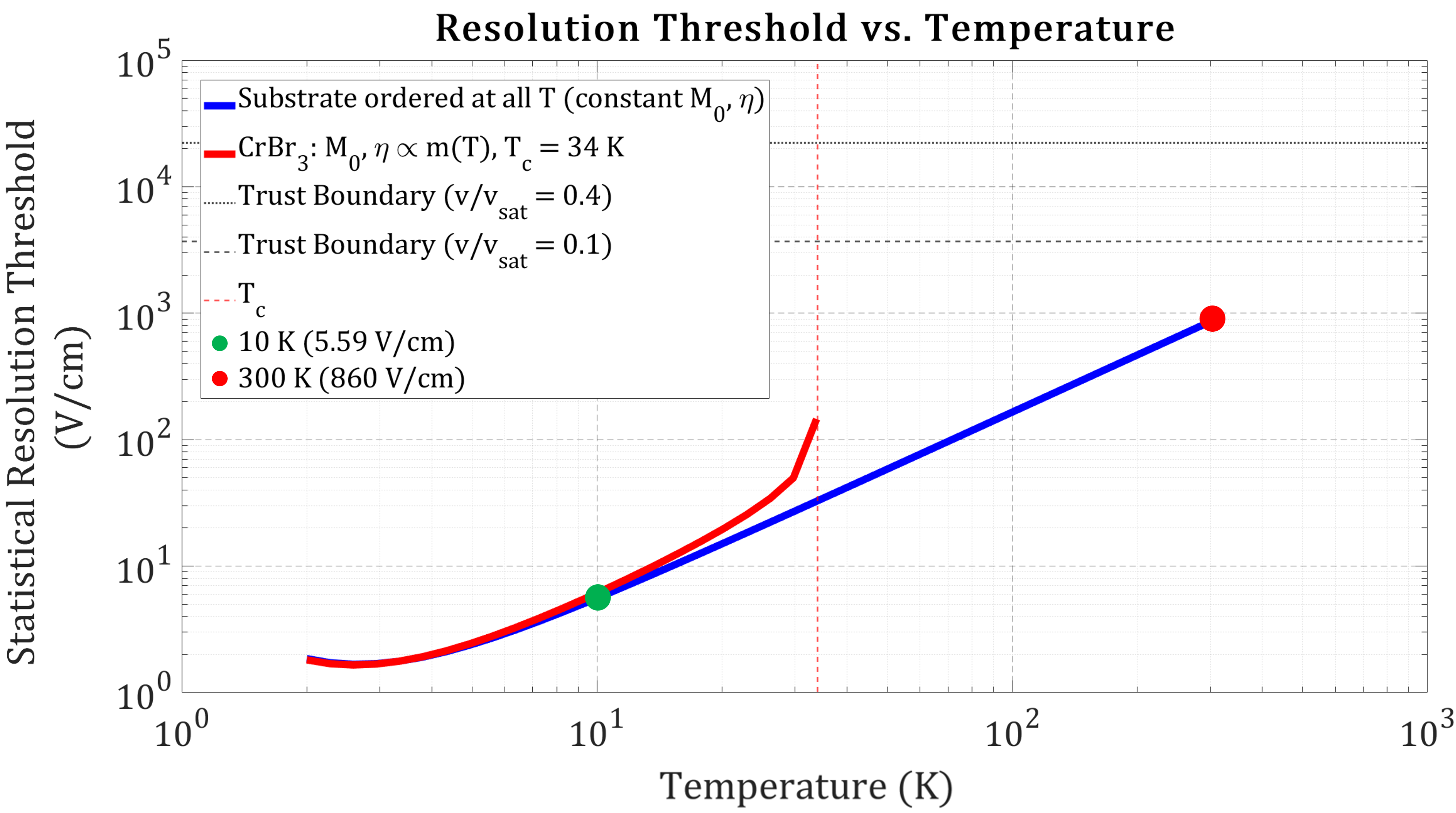}
    \caption{Statistical resolution threshold as a function of temperature. The blue curve is a substrate that stays ordered at every temperature, with $M_0$ and $\eta$ held at their first-principles values. The red curve is $\text{CrBr}_3$, with $M_0$ and $\eta$ scaled by the reduced magnetization $m(T) = (1 - T/T_c)^{0.3}$ and $T_c = 34$ K, and it diverges at $T_c$ (red dashed line). The black dotted and dashed lines are the $v/v_{sat} = 0.4$ and $0.1$ trust boundaries.}
    \label{fig:temperature}
\end{figure}

\begin{figure*}[t]
    \centering
    \includegraphics[width=0.95\linewidth]{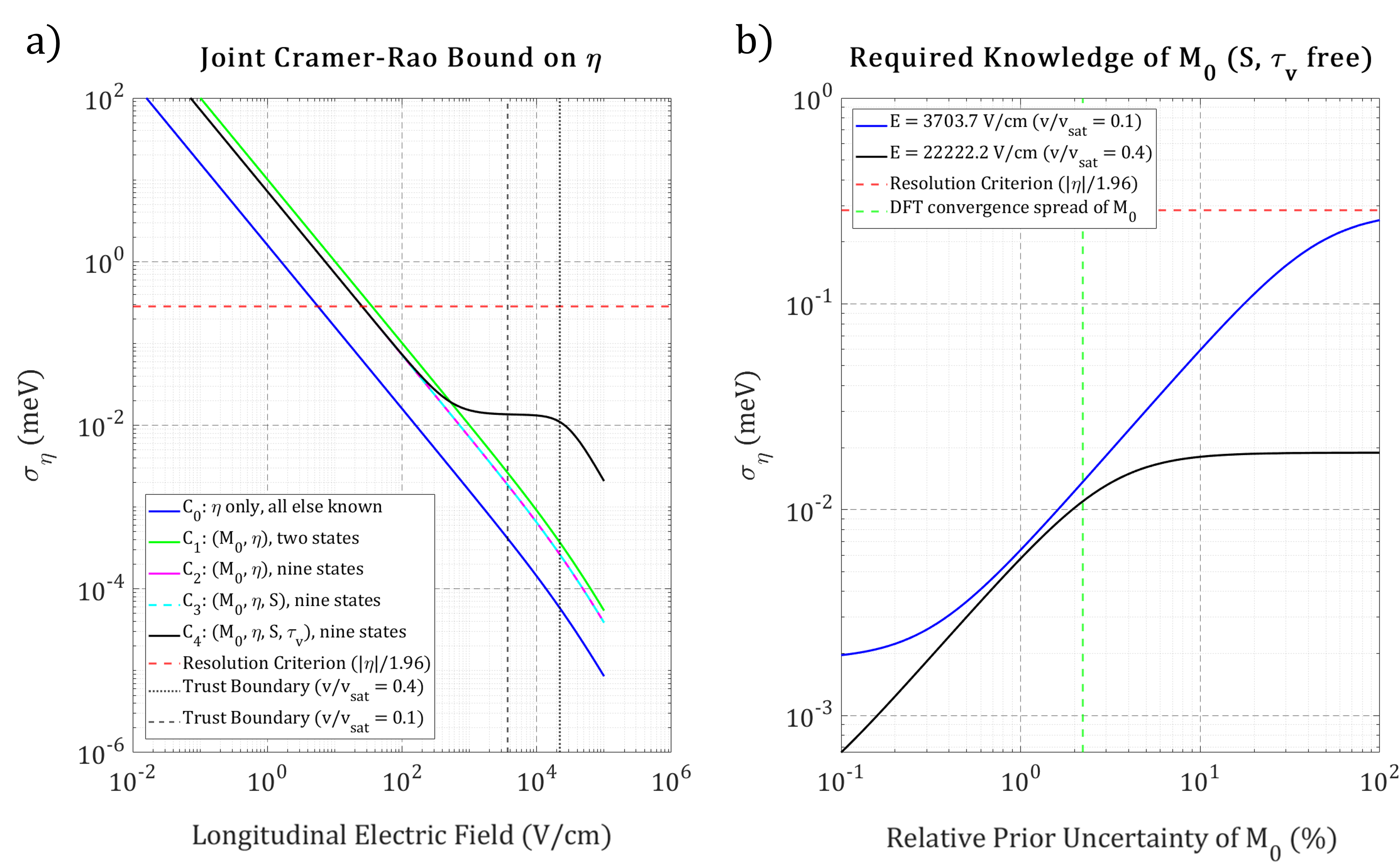}
    \caption{Joint Cram\'er-Rao bound on $\eta$ with nuisance parameters. (a) The bound against the applied field for $\eta$ alone with all other parameters known ($C_0$), for $(M_0, \eta)$ from two and from nine actuation states ($C_1$, $C_2$), with the Hall gain $S$ added ($C_3$, which coincides with $C_2$), and with $\ln\tau_v$ added ($C_4$). The dashed red line is the 95\% resolution criterion $|\eta|/1.96$. (b) The bound of $C_4$ at the two trust boundaries against the relative prior uncertainty of $M_0$, with the 2.2\% convergence spread of Sec. IV.A marked in green.}
    \label{fig:joint}
\end{figure*}

\subsection{Joint Identifiability with Nuisance Parameters}

The bounds of Sec. IV.C treat the zero-strain splitting $M_0$, the Hall gain, and the intervalley scattering time as known. A measurement fixes none of them independently, so this subsection recovers $\eta$ jointly with them and quantifies what each additional unknown costs. The Hall gain $S$ is a single multiplicative factor on $V_H$ that stands for $\beta$, the mobility, and the channel width together. The adjoint sensitivities are extended to the parameter vector $(M_0, \eta, S, \ln\tau_v)$, and the Fisher information is summed over the nine actuation states, each read with 200 independent Hall probes. The zero-strain splitting carries a Gaussian prior of 0.01 meV, the convergence spread of Sec. IV.A, and $\ln\tau_v$ carries a prior of one decade. The Hall gain carries no prior. The calculation treats $M_0$, $S$, and $\tau_v$ as state-independent across the small strain contrast ($\Delta\mathrm{tr}\,\varepsilon \le 0.78 \times 10^{-4}$) and assumes independent noise across states. In a real device, non-magnetic strain effects can be separated from $\eta$ directly. Any strain-linear change in carrier density or mobility shows up in the longitudinal channel conductance $\sigma_{xx}(\varepsilon)$, while strain shifts of the bare $\text{MoS}_2$ band parameters ($\Delta$, $t$, $m^*$) and piezoelectric gating persist unchanged above the Curie temperature ($T > T_c$) and do not flip sign when the substrate magnetization is reversed, allowing them to be subtracted as a baseline. The recovered coupling is the product of $\eta$ and the strain-transfer efficiency from the membrane to the heterostructure, taken here as unity.

Fig.~\ref{fig:joint} and Table S7 of the Supplemental Material \cite{SM} give the results. When $M_0$ is unknown, $\eta$ is identified through the contrast between strain states and not through the absolute strain. The Cram\'er-Rao bound at the 10 percent ceiling of 3,704 V/cm then rises from 413 to 1,870 neV for the nine states and to 2,620 neV for the two extreme states 1 and 9, and the threshold rises from 5.59 to 25.3 and 35.5 V/cm, respectively. Freeing the Hall gain changes neither number, because a gain multiplies a level that is common to all states, and the contrast that carries $\eta$ is orthogonal to it. Freeing $\tau_v$ as well raises the bound to 13.6 $\mu$eV at the 10 percent ceiling, which is 21 times below the 0.286 meV resolution criterion, and to 10.9 $\mu$eV at the 40 percent reference of 22.2 kV/cm, while the threshold stays at 25.3 V/cm.

The bound at the trust boundaries is governed by the prior on $M_0$. In the bulk, the level $A$ of $V_H$ and its strain slope $B$ fix $\eta$ through $\eta = k_BT\,(B/A)\sinh(M_0/k_BT)$, so a relative uncertainty $\delta$ in $M_0$ enters $\eta$ as $1.09\,\delta$ at 10 K, giving 13 $\mu$eV for the 2.2\% convergence spread of $M_0$ (close to the computed 13.6 $\mu$eV at the 10 percent ceiling). This baseline prior is conservative because $M_0$ can also be calibrated independently by magneto-optical spectroscopy at fixed actuator bias, where the static strain shifts the splitting by only 0.2 $\mu$eV. Moreover, Fig.~\ref{fig:joint}(b) shows that the prior is not even required for resolution. With a 100\% prior uncertainty on $M_0$ the bound at the 10 percent ceiling is 0.25 meV, and with no prior at all it saturates at 0.28 meV (below the 0.286 meV criterion) and at 19 $\mu$eV at the 40 percent reference. This saturation occurs because the valley polarization relaxes over a short boundary layer near the contacts, and a strain shift of the splitting alters the amplitude and the relaxation length of that profile with different coefficients, providing information on $M_0$ that is independent of the Hall gain under the detailed-balance rates of Eq.~(11).

\section{Discussion}

This section states the scope of the central claim and the approximations of each stage of the pipeline.

\subsection{Scope of the Central Claim}

The operating window from Section IV is defined for a coupling of a given magnitude. First-principles theory fixes the zero-strain splitting to about 2\% but does not resolve its strain dependence, because the 0.2 $\mu$eV actuator shift lies far below the numerical precision of about 15 $\mu$eV of the splitting differences (Sec. IV.A). The density functional theory calculations therefore bound the coupling at small tensile strain to $|\eta| \lesssim 2$ meV and leave its sign to measurement. The device-level identifiability analysis takes this bound as its starting point and determines the field conditions under which a coupling of the order of the example value of 0.56 meV is statistically resolvable across the operational bias range of a two-dimensional transistor or Hall-bar channel. The contribution of this paper is this general inverse-problem diagnostic and the resulting operating window, which establish before fabrication whether a predicted microscopic coupling can be extracted from a device measurement and what actuation and bias conditions are required to resolve it.

\subsection{Approximations in the Multiscale Model Chain}

The model chain from microscopic quantum physics to the macroscopic device contains several approximations. The parameter $\beta$ is a leading-order Berry curvature evaluated at the $k=0$ band extremum and not a complete Brillouin-zone thermal integral. Thermal averaging over the finite-temperature occupation reduces $\beta$ by a factor that acts as the Hall gain $S$ of Sec. IV.I, where it is treated as unknown. 

The intervalley scattering time is set to $\tau_v = 1$ ns, an order-of-magnitude value from the literature. Recomputing the resolution threshold over two decades of this parameter, $\tau_v \in [0.1, 10]$ ns, leaves it at $5.59$ V/cm. The reason is that the valley diffusion length $L_v = \sqrt{D\tau_v}$ stays below $0.23$ $\mu$m over the swept range, more than two orders of magnitude smaller than the $80$ $\mu$m channel length. The bulk transport observable is set almost entirely by the local, $\tau_v$-independent equilibrium ratio $r(\varepsilon,\eta)$, and the intervalley relaxation kinetics affect only a narrow boundary layer near the contacts. The reduced two-valley system solved here is linear with Dirichlet contacts, and the full drift-diffusion-Poisson system is of the type covered by the existence results for semiconductor device models \cite{markowich1986stationary, gajewski1985existence}.

\subsection{Structural and First-Principles Details}

With PBE, the three stacking registries differ by about 1 meV per supercell, and the energies do not identify a ground state. With the D3 correction, the Chromium-aligned and Bromine-aligned registries lie 5.04 and 13.08 meV above Hollow, and the S--Br gap contracts from 4.26 to 3.32~\AA. The D3 calculations use the PBE in-plane lattice constant, which lies about 0.5\% above the D3 energy minimum of the strained Hollow structure. The valley splitting depends on the registry and on the interlayer separation, with $M_0$ equal to $-0.452$, $-0.248$, and $-0.2104$ meV for the Hollow, Chromium-aligned, and Bromine-aligned registries with D3, and $-0.006$ meV for Hollow with PBE. The coupling $\eta$ is computed for Hollow stacking only. Differences of the splitting between strains are known to about $\pm 0.015$ meV from the convergence tests of Sec. IV.A.

Across the 0 to 4\% strain series, the \texttt{PROCAR} projection weights on the individual $\text{CrBr}_3$ atoms and orbitals for the $\text{MoS}_2$ conduction-band doublet remain at or below the $10^{-3}$ printout resolution of the PAW sphere projection. This makes physical sense, because in second-order perturbation theory a proximity splitting of $|M_0| = 0.45$ meV across a band offset $\Delta_b$ of order $0.1$ to $0.25$ eV requires a total interlayer mixing weight of only $|M_0|/\Delta_b \sim 10^{-3}$ distributed across the substrate orbitals, with part of the overlap residing in the van der Waals gap outside the augmentation spheres.

\subsection{Transport and Device Model}

The transport simulations are isothermal. The 10 percent velocity-saturation convention places the primary ceiling of the operating window at 3,704 V/cm, which corresponds to about 30 V across the channel and lowers the dissipated power by a factor of about 24 relative to the 40 percent reference at 22.2 kV/cm, where about 178 V drop across the channel. The 20 and 40 percent references at 8,333 V/cm and 22.2 kV/cm serve as extended high-field markers.

The transport model is evaluated at 10 K, well below the Curie temperature of $\text{CrBr}_3$, so the substrate is ordered. The thermal reduction of the substrate magnetization rescales $M_0$ and $\eta$ together, and it raises the resolution threshold approximately in inverse proportion to the reduced magnetization (Sec. IV.H). At the contact density used here the Fermi energy of the carrier gas is of the order of a few tenths of a meV, below $k_BT = 0.86$ meV, so the Boltzmann form of the equilibrium ratio and the Einstein relation $D = \mu k_B T / e$ hold to within about 30\% in the equilibrium polarization, which is 0.18 from a Fermi-Dirac evaluation at the contact density against 0.26.

The conversion from three-dimensional to two-dimensional carrier densities uses an equivalent thickness of $d_{thickness} = 0.65$ nm \cite{radisavljevic2011single}. The reduced-dimensionality transport model keeps the problem computationally tractable and is consistent with the reduced-order inverse problem of Section III.

\subsection{Outlook and Future Development}

Four steps would extend this work. A constant-basis strain series with a higher cutoff and a finer mesh will narrow the bound on $\eta$, and a series at the D3 in-plane lattice constant, and for the other stacking registries, will show how much $\eta$ depends on the reference structure. A complete Brillouin-zone Berry curvature integration will replace the leading-order $k=0$ approximation. An ab initio phonon-scattering calculation will replace the phenomenological intervalley scattering time with a computed value. Extending the same pipeline to additional altermagnetic substrate candidates will test the statistical identifiability framework beyond this system.

\section{Conclusion}

This paper gives three important results on the recoverability of strain-tuned proximity valley splitting at the device level. First, structural relaxation lowers the idealized $C_{3v}$ symmetry of the Hollow structure to a chiral $C_3$ point-group symmetry, and a Neumann's-principle analysis shows that the exchange-strain coupling tensor still reduces to a single scalar $\eta$, because the antisymmetric part permitted by the three-fold axis is orthogonal to the symmetric strain tensor. The Bromine-aligned registry, with space group P1, retains independent in-plane components and a singular Fisher information matrix under any single fixed-ratio strain sweep. Second, a continuous partial differential equation-constrained adjoint-state framework formulates the device-level inverse problem directly from the transport equations, yielding statistical resolution bounds through the Fisher information matrix without finite-difference gradients. Third, with continuum elastodynamic and valley-resolved drift-diffusion models, the operating window lies between 5.6 and 3,704 V/cm at 10 K for an example coupling of 0.56 meV when the zero-strain splitting is known. Recovering $\eta$ jointly with the zero-strain splitting, the Hall gain, and the intervalley scattering time moves the lower bound to 25 V/cm and gives a bound of 13.6 $\mu$eV on $\eta$ at the 10 percent ceiling, 21 times below the resolution criterion, with the best actuation design driving the device between its two uniform extremes. Because this range from 25 to 3,704 V/cm spans the full operational drive scale of two-dimensional field-effect transistors, from low-voltage linear readout to the onset of velocity saturation, the coupling can be extracted under standard device operating conditions. For a substrate that remains ordered at 300 K the window spans 860 to 3,704 V/cm, and the joint recovery opens just above the 10 percent ceiling and extends to the 20 and 40 percent references. For $\text{CrBr}_3$ the window closes at the Curie temperature.

The operating window is set by the coupling magnitude. With the dispersion correction the zero-strain valley splitting is $-0.45$ meV, about 75 times its value without it, and the density functional theory calculations bound the exchange-strain coupling at small tensile strain to about 2 meV without fixing its sign. The lower bound of the window scales as $1/|\eta|$. The identifiability analysis therefore serves as a pre-fabrication diagnostic that determines the field conditions under which a target coupling magnitude becomes statistically resolvable against experimental noise before a device is built.

As noted in the Introduction, this framework is not specific to the molybdenum disulfide and chromium tribromide heterostructure. It combines adjoint-state sensitivity analysis, statistical identifiability analysis, and D-optimal experimental design into a general diagnostic tool for two-dimensional transistor and Hall-bar architectures, and it applies to any predicted strain-tunable or field-tunable effect for which a forward model connects the microscopic coupling to a measurable macroscopic observable. Because microscopic material predictions are generated faster than experiments can verify them, applying this inverse framework across candidate topological and magnetic monolayers provides a quantitative route to separate effects that are resolvable in practical devices from those that are not.

\section*{Data Availability}
The density functional theory inputs and outputs (VASP), continuum elastodynamic models (COMSOL), custom adjoint-state optimization scripts (MATLAB), and the data necessary to reproduce the plots within this paper are openly available in Zenodo at \href{https://doi.org/10.5281/zenodo.23646563}{https://doi.org/10.5281/zenodo.23646563}

\begin{acknowledgments}
The author acknowledges the computational resources and simulation frameworks provided by the Department of Electrical and Electronics Engineering at BITS Pilani, Dubai Campus. 
\end{acknowledgments}

\vspace{1em}
\noindent\textit{Note: The consolidated references for both the main text and the Supplementary Material appear at the end of this document, following Supplementary Section~S8.}

\clearpage

\onecolumngrid
\begin{center}
{\large\bfseries Supplementary Material for ``Adjoint-State Identifiability of Predicted Piezo-Tunable Valley Splitting in 2D Magnetic Heterostructures''\par}
\vspace{0.8em}
{\normalsize Suhas Suresh Bharadwaj\textsuperscript{*}\par}
\vspace{0.2em}
{\small\itshape Department of Electrical and Electronics Engineering,\\
BITS Pilani, Dubai Campus, Dubai, United Arab Emirates\par}
\vspace{0.2em}
\end{center}
\vspace{1em}
\twocolumngrid
\begingroup
\endgroup

\setcounter{section}{0}
\setcounter{equation}{0}
\setcounter{figure}{0}
\setcounter{table}{0}
\renewcommand{\thesection}{S\arabic{section}}
\renewcommand{\theequation}{S\arabic{equation}}
\renewcommand{\thefigure}{S\arabic{figure}}
\renewcommand{\thetable}{S\arabic{table}}

\title{Supplementary Material for ``Adjoint-State Identifiability of Predicted Piezo-Tunable Valley Splitting in 2D Magnetic Heterostructures''}

\author{Suhas Suresh Bharadwaj}
\affiliation{Department of Electrical and Electronics Engineering, BITS Pilani, Dubai Campus, Dubai, United Arab Emirates}

\setcounter{section}{0}
\setcounter{equation}{0}
\setcounter{figure}{0}
\setcounter{table}{0}
\renewcommand{\thesection}{S\arabic{section}}
\renewcommand{\theequation}{S\arabic{equation}}
\renewcommand{\thefigure}{S\arabic{figure}}
\renewcommand{\thetable}{S\arabic{table}}

\section{Numerical Discretization and Adjoint Solver Implementation}

This section describes the numerical solver and the checks that were run on it. The solver is a finite-difference code written in MATLAB without external partial differential equation toolboxes, so every step of the adjoint assembly is visible in the code. The diffusion terms use centered finite differences, and the field-driven drift terms use upwind differences, which keeps the scheme stable. The one-dimensional system is solved on a 200-point grid over the 80 $\mu$m channel, which gives a node spacing of 0.4 $\mu$m. The two-dimensional system of Fig. 5 of the main text is discretized on a 120 by 30 grid. The operating map of Fig. 7 reuses the 200-point one-dimensional model at 150 field values from 0.01 to $10^5$ V/cm and 150 channel widths from 1 to 50 $\mu$m. The temperature sweep of Fig. 9 evaluates the Cram\'er-Rao bound at 1 V/cm on the same 200-point grid for temperatures from 2 to 300 K, and the joint analysis of Fig. 10 uses the same grid at 100 field values from 0.01 to $10^5$ V/cm. Both formulations fix the carrier density at the source and drain contacts (Dirichlet conditions). The forward system $A(\eta)u = b$ and the adjoint system $A^T \lambda = \nabla_u J$ are solved directly with the sparse linear solver in MATLAB.

The solver implements the reduced model of Sec. II.F of the main text. Since the Hall angle is small, the field is uniform, and the channel is much longer than the valley diffusion length, the transport problem reduces to two coupled one-dimensional linear drift-diffusion equations for $n_K$ and $n_{K'}$, with the Poisson equation fixing the uniform field and generation and recombination set to zero. In the code, the splitting that enters the intervalley rates is written as $+M(\varepsilon) = M_0 + \eta\,(\varepsilon_{xx} + \varepsilon_{yy})$ rather than $-M(\varepsilon)$. This simply swaps the labels of the $K$ and $K'$ valleys, which flips the sign of the simulated Hall voltage without changing its magnitude, the Fisher information, or any Cram\'er-Rao bound.

The number of Hall probes $N_s$ equals the number of grid nodes, and the probe length equals the node spacing, so that the probes tile the channel. The Johnson noise of a probe is that of the resistance $R = W/(\ell_p \sigma_{xx})$, which grows in proportion to $N_s$, and it exceeds the amplifier noise (at 100 V/cm, $1.05 \times 10^{-7}$ V against $5 \times 10^{-9}$ V for 200 nodes). The information per probe therefore falls as the probe shrinks, and in the Johnson-limited regime the total information depends on the grid only through the contact boundary layers, at the level of 1\%. It matches the information of a single electrode pair of length $L$, whose resistance $W/(L\sigma_{xx})$ is smaller by the factor $N_s$. With 100, 200, and 400 nodes the threshold is 5.61, 5.59, and 5.58 V/cm, and the Cram\'er-Rao bound at the 40 percent trust boundary of 22.2 kV/cm is 57.9, 57.6, and 57.5 neV.

\section{Finite-Difference Gradient Validation}

The adjoint gradient is the quantity the inversion depends on, so it is checked against a finite-difference reference. The adjoint forcing vector includes the dependence on both reciprocal scattering rates, $K \to K'$ and $K' \to K$. Table \ref{tab:gradient_val} gives the check at $\eta = 0$, $E = 1.25$ V/cm, and $W = 10~\mu$m, for one noisy realization of the Hall voltage (random seed 42). The relative error of the central finite difference falls by a factor of 100 for each factor of 10 in step, as expected for a central difference, from $3.4 \times 10^{-3}$ at $10^{-2}$ to $3.4 \times 10^{-9}$ at $10^{-5}$, and reaches $1.8 \times 10^{-10}$ at $10^{-6}$, where rounding limits it. In other words, the two gradients agree to nine digits at the smallest step.

\begin{table}[h]
\caption{\label{tab:gradient_val} Central finite-difference gradient against the adjoint gradient for five step sizes \\}
\centering
\renewcommand{\arraystretch}{1.3}
\begin{tabular}{cccc}
\hline\hline
Step ($\delta \eta$) & FD Gradient & Adjoint Gradient & Relative Error \\
\hline
$10^{-2}$ & $-925.7304$ & $-928.9335$ & $3.4482 \times 10^{-3}$ \\
$10^{-3}$ & $-928.9015$ & $-928.9335$ & $3.4482 \times 10^{-5}$ \\
$10^{-4}$ & $-928.9332$ & $-928.9335$ & $3.4482 \times 10^{-7}$ \\
$10^{-5}$ & $-928.9335$ & $-928.9335$ & $3.4190 \times 10^{-9}$ \\
$10^{-6}$ & $-928.9335$ & $-928.9335$ & $1.8040 \times 10^{-10}$ \\
\hline\hline
\end{tabular}
\end{table}

\section{Intervalley Scattering Time Sensitivity Sweep}

\begin{table}[htbp]
\caption{\label{tab:tau_sweep} Intervalley scattering time sensitivity \\}
\centering
\renewcommand{\arraystretch}{1.3}
\begin{tabular}{ccc}
\hline\hline
$\tau_v$ (ns) & Threshold (V/cm) & $\sigma_\eta$ at 22,222.2 V/cm (neV) \\
\hline
0.1 & 5.59 & 55.8 \\
1.0 & 5.59 & 57.6 \\
10.0 & 5.59 & 86.0 \\
\hline\hline
\end{tabular}
\end{table}

\begin{table*}[t]
\caption{\label{tab:strain_data} Valley splitting of the $\text{MoS}_2$ conduction-band doublet with PBE and with PBE plus the DFT-D3 correction. $\Delta E_{C1}$ and $\Delta E_{C2}$ are the $K$--$K'$ energy differences of the lower and upper spin partner, and $M = (\Delta E_{C1} - \Delta E_{C2})/2$ is the signed splitting. Dashes mark strains that were not computed, and the D3 doublet at $-2$\% is not identified. \\}
\centering
\renewcommand{\arraystretch}{1.2}
\begin{tabular}{ccccccc}
\hline\hline
Strain (\%) & \multicolumn{3}{c}{PBE (meV)} & \multicolumn{3}{c}{PBE + D3 (meV)} \\
 & $\Delta E_{C1}$ & $\Delta E_{C2}$ & $M$ & $\Delta E_{C1}$ & $\Delta E_{C2}$ & $M$ \\
\hline
$-3$ & $+0.0470$ & $-0.0480$ & $+0.0475$ & -- & -- & -- \\
$-2.5$ & $+0.0430$ & $-0.0440$ & $+0.0435$ & -- & -- & -- \\
$-2$ & $+0.1820$ & $-0.1020$ & $+0.1420$ & -- & -- & -- \\
$-1.5$ & $-0.0840$ & $+0.0580$ & $-0.0710$ & $-0.1240$ & $+0.4580$ & $-0.2910$ \\
$-1$ & $-0.0110$ & $+0.0200$ & $-0.0155$ & $-0.5380$ & $+0.6780$ & $-0.6080$ \\
$-0.5$ & -- & -- & -- & $-0.4450$ & $+0.2470$ & $-0.3460$ \\
$-0.25$ & -- & -- & -- & $-0.4450$ & $+0.3860$ & $-0.4155$ \\
0 & $-0.0070$ & $+0.0050$ & $-0.0060$ & $-0.4690$ & $+0.4350$ & $-0.4520$ \\
$+0.25$ & -- & -- & -- & $-0.4680$ & $+0.4420$ & $-0.4550$ \\
$+0.5$ & -- & -- & -- & $-0.4710$ & $+0.4510$ & $-0.4610$ \\
$+0.75$ & -- & -- & -- & $-0.4570$ & $+0.4370$ & $-0.4470$ \\
$+1$ & $+0.0030$ & $-0.0040$ & $+0.0035$ & $-0.4510$ & $+0.4330$ & $-0.4420$ \\
$+2$ & $-0.0120$ & $+0.0130$ & $-0.0125$ & $-0.4970$ & $+0.4920$ & $-0.4945$ \\
$+3$ & $-0.0310$ & $+0.0310$ & $-0.0310$ & $-0.5740$ & $+0.5670$ & $-0.5705$ \\
$+4$ & $-0.0190$ & $+0.0200$ & $-0.0195$ & $-0.6440$ & $+0.6250$ & $-0.6345$ \\
\hline\hline
\end{tabular}
\end{table*}

The intervalley scattering time $\tau_v$ is an input to the valley-resolved transport model. This section tests how much the operating window depends on it. Table \ref{tab:tau_sweep} gives the statistical resolution threshold, and the Cram\'er-Rao bound at the 40 percent trust boundary of 22.2 kV/cm, recomputed over two decades of $\tau_v$. The threshold stays at 5.59 V/cm over this range. The reason is the valley diffusion length $L_v = \sqrt{D \tau_v}$, which is about 0.23 $\mu$m at 10 ns, more than two orders of magnitude below the 80 $\mu$m channel length. The bulk transport observable is therefore set almost entirely by the local equilibrium ratio, and the intervalley relaxation kinetics affect only a narrow boundary layer near the contacts. At the 40 percent trust boundary of 22.2 kV/cm the drift length $v_d \tau_v$ reaches 80 $\mu$m for $\tau_v = 10$ ns, and the Cram\'er-Rao bound there changes from 55.8 to 86.0 neV across the sweep, a factor of 1.5.

\section{Exchange-Strain Regression and Statistics}

This section lists the strain data and the fits behind the exchange-strain coupling $\eta$. The coupling is extracted from strain series in which the ions are relaxed at each strain (Sec. S5), with and without the DFT-D3 correction. At each strain, the splitting is the energy difference between the $K$ and $K'$ points of each partner of the lowest $\text{MoS}_2$ conduction-band spin doublet, and $M$ is half the difference of the two partner values, $M = (\Delta E_{C1} - \Delta E_{C2})/2$. One frozen rule finds this doublet in all calculations. It takes the lowest conduction pair with a $\text{MoS}_2$ projection weight above 0.78 at both $K$ and $K'$ and opposite-sign $m_z$ at $K$ and at $K'$. In other words, the two bands must be mostly $\text{MoS}_2$ and must be spin partners. Table \ref{tab:strain_data} lists the splitting of each spin partner and the signed splitting $M$. The energies come from the \texttt{EIGENVAL} files, which have a resolution of 1 $\mu$eV, so the listed splittings are uncertain by about 0.001 meV from rounding. The shipped extraction script, which reads the 8-decimal energies in the \texttt{PROCAR} files, reproduces every value in the table to within 0.001 meV. The regressions below use the table values. Refitting with the \texttt{PROCAR} values changes $\eta$ by at most 0.07 meV and leaves the conclusions unchanged.

With D3 the doublet is bands 163 and 164 from $-0.5$ to $+1$\% strain, bands 162 and 163 at $+2$ to $+4$\%, bands 165 and 166 at $-1$\%, and bands 166 and 167 at $-1.5$\%. At $-2$\% it is not identified. The upper partner at $K'$ is spread over bands 170 to 172, with $\text{MoS}_2$ weights of 0.70, 0.66, and 0.66, so no single band passes the 0.78 threshold.

\begin{table}[htbp]
\caption{\label{tab:bands_d3} D3 series with the distance in meV, at $K$, from the lower partner of the doublet to the nearest band with $\text{CrBr}_3$ weight above 0.1, and to the nearest band with both $\text{MoS}_2$ and $\text{CrBr}_3$ weight above 0.1. Positive values lie above the doublet. \\}
\centering
\renewcommand{\arraystretch}{1.2}
\begin{tabular}{ccc}
\hline\hline
Strain (\%) & Nearest $\text{CrBr}_3$ band & Nearest mixed band \\
\hline
$-1.5$ & $-23.1$ & $-23.1$ \\
$-1$ & $-62.7$ & $+68.9$ \\
$-0.5$ & $+22.4$ & $+158.3$ \\
$-0.25$ & $+65.6$ & $+202.5$ \\
0 & $+110.6$ & $+247.1$ \\
$+0.25$ & $+151.7$ & $+288.7$ \\
$+0.5$ & $-121.9$ & $+331.2$ \\
$+0.75$ & $-91.8$ & $+373.8$ \\
$+1$ & $-62.3$ & $+414.8$ \\
$+2$ & $+54.0$ & $+569.1$ \\
$+3$ & $-107.8$ & $+703.7$ \\
$+4$ & $-18.1$ & $+815.3$ \\
\hline\hline
\end{tabular}
\end{table}

The primary fit uses the D3 points from 0 to $+1$\% in steps of 0.25\%. Ordinary least squares through these five points gives a slope of $+0.0112$ meV per percent strain with a standard error of 0.0085 meV per percent (three degrees of freedom). Converting to fractional strain and dividing by the factor of two from the equal-biaxial protocol gives $\eta = +0.56$ meV with a 95 percent confidence interval of $[-0.79, +1.91]$ meV (Sec. II.D of the main text). This 0 to $+1$\% tensile window is the one relevant to the AlN piezoelectric actuator, which applies a purely tensile bias trace of $3.34$ to $4.12 \times 10^{-4}$, and it avoids the band crossings at $-0.64$\% and $-1.4$\% strain that perturb the compressive branch (Table \ref{tab:bands_d3}).

The slope changes in sign and size between windows, because $M$ is not linear in strain over a few percent. Table \ref{tab:windows} lists the fits over six windows, with $\eta$ between $-5.39$ and $+0.56$ meV. In other words, the value of $\eta$ depends on which strain window is fitted, and the tensile window from 0 to $+1$\% is the one that matches the actuator. No bootstrap is used for D3, since the residuals of a line are set by the curvature of $M$ and not by sampling noise. The fits also carry the numerical spread of the strain differences. The difference $M(+1\%) - M(0\%)$ ranges from $-0.013$ to $+0.017$ meV across the five settings of Sec. S5, against $+0.010$ meV at the baseline, so these fits bound $|\eta|$ and do not fix its sign.

\begin{table}[h]
\caption{\label{tab:windows} Linear fits of the D3 splitting $M$ over different strain windows. The 95 percent intervals use a Student $t$ distribution with $n-2$ degrees of freedom, which is very wide for $n = 3$. \\}
\centering
\renewcommand{\arraystretch}{1.3}
\setlength{\tabcolsep}{3.5pt}
\begin{tabular}{ccccc}
\hline\hline
Window & $n$ & Slope & $\eta$ & 95\% interval \\
(\%) & & (meV/\%) & (meV) & (meV) \\
\hline
0 to $+1$ (primary) & 5 & $+0.0112$ & $+0.56$ & $[-0.79, +1.91]$ \\
0 to $+0.5$ & 3 & $-0.018$ & $-0.90$ & $[-3.10, +1.30]$ \\
$-0.25$ to $+0.25$ & 3 & $-0.079$ & $-3.95$ & $[-28.5, +20.6]$ \\
$-0.5$ to $+0.5$ & 5 & $-0.108$ & $-5.39$ & $[-10.6, -0.21]$ \\
$-0.25$ to $+1$ & 6 & $-0.0141$ & $-0.71$ & $[-2.88, +1.46]$ \\
0 to $+4$ & 8 & $-0.0469$ & $-2.34$ & $[-3.11, -1.58]$ \\
\hline\hline
\end{tabular}
\end{table}

At 10 K the thermal occupation is confined almost entirely to the lower spin partner. An ordinary least-squares fit of $\Delta E_{C1}$ alone over the same five points gives a slope of $+0.0188 \pm 0.0067$ meV per percent strain (three degrees of freedom), which gives $\eta = +0.94$ meV with a 95 percent confidence interval of $[-0.12, +2.00]$ meV.

The PBE series is the weak-coupling comparison. Its zero-strain splitting is smaller than the D3 value by a factor of 75. A line through 0 to $+4$\% has a slope of $-0.00615 \pm 0.00321$ meV per percent ($R^2 = 0.550$, $p = 0.15$) and gives $\eta = -0.31$ meV with a 95 percent interval of $[-0.82, +0.20]$ meV. Including $-1$\% gives $\eta = -0.16$ meV with $[-0.54, +0.22]$ meV, and including $-1.5$ and $-2$\% gives $\eta = -0.52$ meV with $[-1.84, +0.80]$ meV. All three intervals include zero.

With PBE the mixed band crosses the doublet between $-1.5$ and $-2$\% strain, from 14.0 meV above to 12.2 meV below the lower partner at $K$. Over the same strains the two partner splittings differ in magnitude by 26 and 80 $\mu$eV, and $M$ changes sign from $-0.071$ to $+0.142$ meV. At $-2.5$ and $-3$\% the doublet is still the lowest $\text{MoS}_2$ pair, with a spin splitting of 1.0 and 0.035 meV.

\section{Density Functional Theory Computational Parameters}

This section lists the settings of the density functional theory calculations. They use VASP 6.5.1 with the projector augmented-wave (PAW) method and the Perdew-Burke-Ernzerhof (PBE) exchange-correlation functional. The potentials are Mo\_sv, S, Cr, and Br from the PAW\_PBE set. No Hubbard $U$ is applied. The PBE series has no van der Waals correction. The D3 series adds the zero-damping DFT-D3 correction (IVDW = 11) in the relaxation, collinear, and spin-orbit steps, at the PBE in-plane lattice constant. The supercell has a lattice constant of 6.3678 \AA\ (twice the MoS$_2$ value of 3.184 \AA), a cell height of 35 \AA, and 20 atoms (4 Mo, 8 S, 2 Cr, 6 Br) with 158 valence electrons.

The plane-wave cutoff is 600 eV, the self-consistent calculations of the strain series use a $\Gamma$-centered $6 \times 6 \times 1$ Brillouin zone mesh, and the smearing is Gaussian with a width of 0.05 eV. The electronic convergence threshold is $10^{-7}$ eV. A collinear calculation with the total moment fixed at 6.00 $\mu_B$ first converges the charge density. The noncollinear spin-orbit calculation then starts from it (ISTART = 0, ICHARG = 1, NELMDL = $-8$, AMIX = 0.05, AMIX\_MAG = 0.10, no symmetry), with the spin axis along the out-of-plane direction and an initial moment of $3~\mu_B$ on each Cr atom. The converged noncollinear calculations give a total magnetization of 6.00 $\mu_B$ along $z$. The valley splitting is read from this calculation, in which $K$ and $K'$, at $(1/3, 1/3, 0)$ and $(-1/3, -1/3, 0)$ in reciprocal coordinates, are points of the $6 \times 6 \times 1$ mesh.

The stacking comparison uses collinear, spin-polarized relaxations at fixed cell shape, with the total moment fixed at 6.00 $\mu_B$, a plane-wave cutoff of 500 eV, a $\Gamma$-centered $6 \times 6 \times 1$ mesh, the conjugate-gradient algorithm, a force threshold of 5 meV/\AA, and at most 150 ionic steps. All three runs reach the threshold, with maximum forces of 4.0 to 4.8 meV/\AA\ and free energies of $-122.55675$ eV (Hollow), $-122.55611$ eV (Chromium-aligned), and $-122.55573$ eV (Bromine-aligned). The vertical gap between the lowest Br plane and the highest S plane is 4.26, 4.24, and 4.26 \AA\ for the Hollow, Chromium-aligned, and Bromine-aligned structures. At 600 eV with PBE, the Chromium-aligned and Bromine-aligned registries lie 0.70 and 1.06 meV above Hollow, with gaps of 4.244 and 4.262 \AA\ and forces of 3.97 and 4.53 meV/\AA.

The D3 comparison uses the relaxation and electronic settings of the strain series below. It gives free energies of $-126.29776$ eV (Hollow), $-126.29272$ eV (Chromium-aligned), and $-126.28468$ eV (Bromine-aligned), so the last two lie 5.04 and 13.08 meV above Hollow. The S--Br gaps are 3.322, 3.343, and 3.360 \AA, and the maximum forces are 4.85, 4.81, and 4.82 meV/\AA. The Bromine-aligned D3 relaxation reached 9.6 meV/\AA\ in 58 conjugate-gradient steps and was finished in 13 steps with IBRION = 1 and POTIM = 0.25.

The convergence of the splitting was tested on the D3 structure at 0 and $+1$\% strain by repeating the spin-orbit calculation with the geometry fixed. The values come from bands 163 and 164 at $K$ and $K'$, read from the \texttt{PROCAR} files. At 0\% strain, $M_0$ is $-0.4256$, $-0.4521$, $-0.4553$, and $-0.4448$ meV at cutoffs of 500, 600, 700, and 800 eV on the $6 \times 6 \times 1$ mesh, and $-0.4543$ meV on a $9 \times 9 \times 1$ mesh at 600 eV. The values of $\Delta E_{C1}$ and $\Delta E_{C2}$ at 0\% strain are $-0.4433$ and $+0.4078$ meV at 500 eV, $-0.4698$ and $+0.4344$ meV at 600 eV, $-0.4731$ and $+0.4376$ meV at 700 eV, and $-0.4721$ and $+0.4365$ meV on the $9 \times 9 \times 1$ mesh. At $+1$\% strain, $M$ is $-0.4420$ meV at 600 eV ($\Delta E_{C1} = -0.4508$ meV and $\Delta E_{C2} = +0.4331$ meV), $-0.4595$ meV at 700 eV ($-0.4684$ and $+0.4506$ meV), and $-0.4373$ meV on the finer mesh ($-0.4461$ and $+0.4285$ meV). The difference $M(+1\%) - M(0\%)$ is $+0.0101$, $-0.0042$, and $+0.0170$ meV for these three settings.

At a fixed cutoff the number of plane waves changes with the cell size. A second set of calculations keeps the basis the same size by scaling the cutoff as $(1+\varepsilon)^{-4/3}$. From a base cutoff of 600 eV, $M$ is $-0.4521$, $-0.4585$, and $-0.4468$ meV at 0, $+0.5$, and $+1$\% strain, with cutoffs of 600.0, 596.0, and 592.1 eV. From a base cutoff of 800 eV, $M$ is $-0.4448$, $-0.4464$, and $-0.4580$ meV, with cutoffs of 800.0, 794.7, and 789.4 eV. The difference $M(+1\%) - M(0\%)$ is $+0.0053$ meV from the 600 eV base and $-0.0132$ meV from the 800 eV base. Across all five settings it ranges from $-0.0132$ to $+0.0170$ meV, and the corresponding two-point estimates of $\eta$ range from $-0.66$ to $+0.85$ meV.

The relaxed structures at 0\% strain were analyzed with spglib at tolerances from $10^{-6}$ to $10^{-2}$. The Hollow and Chromium-aligned structures have space group P3 (No. 143) with PBE and with D3 at every tolerance, and the Bromine-aligned structures have space group P1 (No. 1).

The PBE strain series scales the in-plane lattice vectors of the relaxed Hollow structure by $-3$, $-2.5$, $-2$, $-1.5$, $-1$, 0, $+1$, $+2$, $+3$, and $+4$\% about 6.36777~\AA\ and keeps the cell height at 35~\AA. The ions are relaxed at each strain with the conjugate-gradient algorithm at 600 eV on the $6 \times 6 \times 1$ mesh, with the total moment fixed at 6.00 $\mu_B$, a force threshold of 5 meV/\AA, and at most 150 ionic steps. Each relaxation is followed by a collinear self-consistent calculation and a noncollinear spin-orbit calculation at the same cutoff and mesh. All ten strains reach the threshold, with final maximum forces of 3.53, 3.38, 3.57, 3.92, 3.46, 3.49, 4.67, 4.61, 4.45, and 3.58 meV/\AA\ from $-3$ to $+4$\%. The vertical S--Br gap is 4.203, 4.213, 4.220, 4.251, 4.259, 4.263, 4.277, 4.314, 4.313, and 4.384 \AA\ over the same strains.

In the PBE comparison series, the 0\% and $-1$\% points reached a force floor near 16 meV/\AA\ with real-space projectors (LREAL = Auto) and were relaxed with reciprocal-space projectors (LREAL = .FALSE.) and EDIFF = $10^{-6}$ eV to reach the 5 meV/\AA\ threshold. The $-1.5$, $-2.5$, and $-3$\% points were relaxed with LREAL = .FALSE. and EDIFF = $10^{-7}$ eV. The other PBE strains use LREAL = Auto and EDIFF = $10^{-7}$ eV, and the electronic stages use LREAL = Auto at all PBE strains. Total energies at different PBE strains are not compared.

The D3 series scales the same lattice vectors by $-2$, $-1.5$, $-1$, $-0.5$, $-0.25$, 0, $+0.25$, $+0.5$, $+0.75$, $+1$, $+2$, $+3$, and $+4$\% and uses one relaxation protocol at every strain. Every D3 relaxation uses LREAL = .FALSE., EDIFF = $10^{-7}$ eV, and IVDW = 11, followed by the same collinear and spin-orbit steps with IVDW = 11. All D3 strains reach the threshold, with final maximum forces of 2.51, 4.24, 4.22, 3.05, 2.85, 4.85, 4.78, 4.85, 2.06, 3.48, 4.07, 4.51, and 4.82 meV/\AA\ and S--Br gaps of 3.348, 3.333, 3.331, 3.338, 3.336, 3.322, 3.327, 3.332, 3.335, 3.339, 3.331, 3.328, and 3.307 \AA\ over the same strains. The D3 total energy is lowest near $-0.5$\% strain ($-11.5$ meV relative to 0\%). The PBE in-plane lattice constant therefore lies about 0.5\% above the D3 minimum.

\section{Aluminum Nitride Actuator Model Specification}

This section gives the material constants and the nine actuation states of the continuum model. The continuum elastodynamic simulations use standard aluminum nitride properties, consistent with the measured elastic and piezoelectric constants of wurtzite AlN reported by Tsubouchi and Mikoshiba \cite{alnConstants}, as implemented in the COMSOL Material Library. The density is 3300 kg/m$^3$. The elasticity tensor $c^E$ in stress-charge form, in GPa and Voigt notation, is
\begin{equation}
c^E = \begin{bmatrix} 
410 & 149 & 99 & 0 & 0 & 0 \\ 
149 & 410 & 99 & 0 & 0 & 0 \\ 
99 & 99 & 389 & 0 & 0 & 0 \\ 
0 & 0 & 0 & 125 & 0 & 0 \\ 
0 & 0 & 0 & 0 & 125 & 0 \\ 
0 & 0 & 0 & 0 & 0 & 130.5 
\end{bmatrix}
\end{equation}
The piezoelectric coupling tensor $e$ in stress-charge form, in C/m$^2$, is
\begin{equation}
e = \begin{bmatrix} 
0 & 0 & 0 & 0 & -0.48 & 0 \\ 
0 & 0 & 0 & -0.48 & 0 & 0 \\ 
-0.58 & -0.58 & 1.55 & 0 & 0 & 0 
\end{bmatrix}
\end{equation}
The relative permittivity tensor $\epsilon_r^S$ is isotropic with a diagonal value of 9. The simulated membrane cross-section is 200 $\mu$m long and 1 $\mu$m thick, and a nine-state voltage sweep is applied across the electrode geometry. Within the 80 $\mu$m active region, the strain trace is uniform to within 0.01\% of its mean for each state. It spans $3.34 \times 10^{-4}$ to $4.12 \times 10^{-4}$ across the nine states (Table \ref{tab:strain_states}), which is the strain range available to the actuator.

\begin{table}[h]
\caption{\label{tab:strain_states} Strain trace and strain ratio for the nine actuation states, averaged over the active region at $y = 0.5~\mu$m. The last column is the squared ratio of each trace to that of state 9. \\}
\centering
\renewcommand{\arraystretch}{1.3}
\begin{tabular}{cccccc}
\hline\hline
State & $V_{\mathrm{inner}}$ (V) & $V_{\mathrm{outer}}$ (V) & Trace ($10^{-4}$) & $\varepsilon_{yy}/\varepsilon_{xx}$ & $(\mathrm{tr}/\mathrm{tr}_9)^2$ \\
\hline
1 & $-10$ & $-10$ & 4.1231 & $-0.1747$ & 1.523 \\
2 & $-10$ & $0$ & 4.0372 & $-0.1729$ & 1.460 \\
3 & $-10$ & $+10$ & 3.9513 & $-0.1709$ & 1.399 \\
4 & $0$ & $-10$ & 3.8180 & $-0.2545$ & 1.306 \\
5 & $0$ & $0$ & 3.7320 & $-0.2545$ & 1.248 \\
6 & $0$ & $+10$ & 3.6461 & $-0.2545$ & 1.191 \\
7 & $+10$ & $-10$ & 3.5128 & $-0.3304$ & 1.106 \\
8 & $+10$ & $0$ & 3.4268 & $-0.3322$ & 1.052 \\
9 & $+10$ & $+10$ & 3.3409 & $-0.3339$ & 1.000 \\
\hline\hline
\end{tabular}
\end{table}

\section{D-Optimal Experimental Design Numerical Detail}

This section gives the two-dimensional check of the D-optimal comparison of Sec. IV.E of the main text, for known $M_0$. The check uses the two-dimensional forward model on the 120 by 30 grid, at $E = 1.25$ V/cm and $W = 1~\mu$m. Both contacts hold the carrier density fixed at $n_{contact} = 10^{18}\text{ cm}^{-3}$, as in the one-dimensional model, so the two discretizations share the same boundary conditions. The effective hopping parameter is $t = 1.1601$ eV. The score of a state is the squared sum of the adjoint sensitivity kernel, $(\sum K)^2$, which is the squared sensitivity of the Hall voltage to $\eta$. It is not divided by the noise variance, so only ratios between states carry meaning.

The score is $3.5041 \times 10^{-13}$ for the uniform state and $3.8738 \times 10^{-13}$ for the split-electrode state, a ratio of 1.1055 that matches the 10.55\% increase in Sec. IV.E of the main text. This ratio equals $(3.5128/3.3409)^2$, the squared ratio of the two strain traces. In the small-signal regime the sensitivity is linear in the strain trace, so the score scales with the squared trace. The same scaling gives the relative scores of the other seven states in Table \ref{tab:strain_states}.

\section{Symmetry-Collapse Derivation}

This section gives the derivation behind the illustration of Sec. III.F of the main text. The dependence of identifiability on the point-group symmetry is illustrated with a lower-symmetry registry such as the Bromine-aligned structure (space group P1), which retains independent in-plane coupling constants, represented here by $\eta_{xx}$ and $\eta_{yy}$. The strain ratio is fixed at $\varepsilon_{yy} = -\nu\,\varepsilon_{xx}$ with $\nu = 0.27$, a value of the order of the Poisson ratio of monolayer MoS$_2$. This is the ratio of the illustration only, and it differs from the actuator ratio used for the operating window in the main text. The fixed ratio gives $K_{yy} = -\nu K_{xx}$ at every node, so the two-by-two Fisher information matrix is
\begin{equation}
F = F_{11}\begin{pmatrix} 1 & -\nu \\ -\nu & \nu^2 \end{pmatrix},
\end{equation}
where $F_{11} = \sum_i (K_{xx,i}/\sigma_V)^2$. Its determinant is $F_{11}F_{22} - F_{12}^2 = \nu^2 F_{11}^2 - \nu^2 F_{11}^2 = 0$. Intuitively, the data depend only on the combination $\eta_{xx} - \nu\,\eta_{yy}$, so a single fixed-ratio actuation state cannot separate the two constants.

In the computed illustration at $E = 1000$ V/cm and $W = 10~\mu$m, $F_{11} \approx 2.50 \times 10^{11}$, $F_{22} \approx 1.82 \times 10^{10}$ and $F_{12} \approx -6.76 \times 10^{10}$ eV$^{-2}$, and the determinant evaluates to $-9.5 \times 10^{5}$ against $F_{11}F_{22} \approx 4.6 \times 10^{21}$, which is zero to floating-point precision. The eigenvalues of $F$ are $2.69 \times 10^{11}$ and $-3.8 \times 10^{-6}$ eV$^{-2}$, a ratio of magnitude $1.4 \times 10^{-17}$. The confirmed $C_3$ symmetry sets $\eta_{xx} = \eta_{yy} = \eta$. This leaves one parameter with $F = (1-\nu)^2 F_{11}$ and $\sigma_\eta = 2.7~\mu$eV at this field, so the rank deficiency does not arise.

\section{Joint Identifiability Numerical Detail}

\begin{table*}[t]
\caption{\label{tab:joint} Joint Cram\'er-Rao bound on $\eta$ for five sets of unknown parameters. The prior on $M_0$ is 0.01 meV, the prior on $\ln\tau_v$ is one decade, and the Hall gain $S$ carries no prior. The two fields are the 10 and 40 percent trust boundaries. \\}
\centering
\renewcommand{\arraystretch}{1.3}
\begin{tabular}{lccc}
\hline\hline
Configuration & Threshold (V/cm) & $\sigma_\eta$ at 3,703.7 V/cm (neV) & $\sigma_\eta$ at 22,222.2 V/cm (neV) \\
\hline
$C_0$: $\eta$ only, state 7 & 5.59 & 413 & 57.6 \\
$C_1$: $(M_0, \eta)$, states 1 and 9 & 35.5 & 2,620 & 366 \\
$C_2$: $(M_0, \eta)$, nine states & 25.3 & 1,870 & 261 \\
$C_3$: $(M_0, \eta, S)$, nine states & 25.3 & 1,870 & 261 \\
$C_4$: $(M_0, \eta, S, \ln\tau_v)$, nine states & 25.3 & 13,600 & 10,900 \\
\hline\hline
\end{tabular}
\end{table*}

\begin{table*}[htbp]
\caption{\label{tab:joint300} Joint Cram\'er-Rao bound on $\eta$ at 300 K for a substrate that stays ordered, for the five sets of unknown parameters of Table \ref{tab:joint}. The 95\% resolution criterion is $|\eta|/1.96 = 286~\mu$eV. The single-parameter threshold of 850 V/cm includes the field dependence of the mobility at the threshold, and it differs by about 1\% from the 860 V/cm of the temperature sweep in the main text, which scales the bound from 1 V/cm. \\}
\centering
\renewcommand{\arraystretch}{1.3}
\begin{tabular}{lccc}
\hline\hline
Configuration & Threshold (V/cm) & $\sigma_\eta$ at 3,703.7 V/cm ($\mu$eV) & $\sigma_\eta$ at 22,222.2 V/cm ($\mu$eV) \\
\hline
$C_0$: $\eta$ only, state 7 & 850 & 63.4 & 8.88 \\
$C_1$: $(M_0, \eta)$, states 1 and 9 & 5,140 & 403 & 56.4 \\
$C_2$: $(M_0, \eta)$, nine states & 3,720 & 287 & 40.2 \\
$C_3$: $(M_0, \eta, S)$, nine states & 3,720 & 287 & 42.0 \\
$C_4$: $(M_0, \eta, S, \ln\tau_v)$, nine states & 3,720 & 287 & 42.0 \\
\hline\hline
\end{tabular}
\end{table*}

This section gives the numerical detail behind Sec. IV.I of the main text. The calculation uses the one-dimensional forward model on the 200-point grid at 100 field values from 0.01 to $10^5$ V/cm and the strain traces of the nine actuation states of Table \ref{tab:strain_states}, evaluated along the channel at $y = 0.5~\mu$m. The mean traces computed from the exported finite-element solution agree with Table \ref{tab:strain_states} to four digits. For each state the adjoint sensitivities of the Hall voltage to $M_0$, $\eta$, the Hall gain $S$, and $\ln\tau_v$ are computed at the first-principles parameter values, and the Fisher information of the state is the sum over the 200 sensors of the squared sensitivities divided by the noise variance of that state, which is that of Sec. II.F of the main text for Hall probes of length equal to the node spacing. The derivative with respect to $\ln\tau_v$ uses the fact that the scattering part of the system matrix scales as $1/\tau_v$, so that $\partial A/\partial \ln\tau_v = -2\,[A(\tau_v) - A(2\tau_v)]$. The states are treated as independent measurements, and the total information is the sum over states. The prior on $M_0$ is Gaussian with a standard deviation of 0.01 meV, the convergence spread of Sec. IV.A of the main text, and the prior on $\ln\tau_v$ has a standard deviation of $\ln 10$. The Hall gain carries no prior. The Cram\'er-Rao bound on $\eta$ is the corresponding diagonal element of the inverse of the total information matrix. Table \ref{tab:joint} lists the five configurations, and configuration $C_0$ reproduces the single-state bound of Sec. IV.C of the main text.

The D-optimal comparison of Sec. IV.E scores each pair of states by the determinant of the two-by-two Fisher information matrix for $(M_0, \eta)$ at 100 V/cm, and the ratio of the determinants is independent of the field in the small-signal regime. The pairs (1, 9), (1, 8) and (2, 9), and (1, 7) and (2, 8) have strain contrasts of $0.78$, $0.70$, and $0.61 \times 10^{-4}$ and scores of 1.00, 0.79, and 0.61, equal to the squared ratios of the contrasts to that of the best pair.

Table \ref{tab:joint300} gives the same five configurations at 300 K for a substrate that stays ordered, where the equilibrium polarization is 0.9\%. Once $M_0$ is constrained by its prior ($C_2$), freeing the Hall gain ($C_3$) and the intervalley scattering time ($C_4$) adds almost no further uncertainty at 300 K, as $C_2$, $C_3$, and $C_4$ differ by less than 5\% at the 22.2 kV/cm reference. The prior on $M_0$ itself still matters at 300 K, because at the 10 percent ceiling the joint bound sits right at the resolution criterion ($\eta$ is resolved there only if $M_0$ is known to better than 0.1\%), whereas at the 22.2 kV/cm reference an uncertainty of up to 54\% in $M_0$ is tolerated.

\section{Recovery Test on Synthetic Data}

This section tests that the inversion attains the Cram\'er-Rao bound. Synthetic Hall voltages are generated for the nine actuation states of Table \ref{tab:strain_states} at $E = 100$ V/cm with the first-principles parameters ($M_0 = -0.452$ meV, $\eta = +0.56$ meV) and Gaussian noise with the variance of Sec. II.F of the main text, for 200 Hall probes per state. 

\begin{figure}[htbp]
    \centering
    \includegraphics[width=\columnwidth]{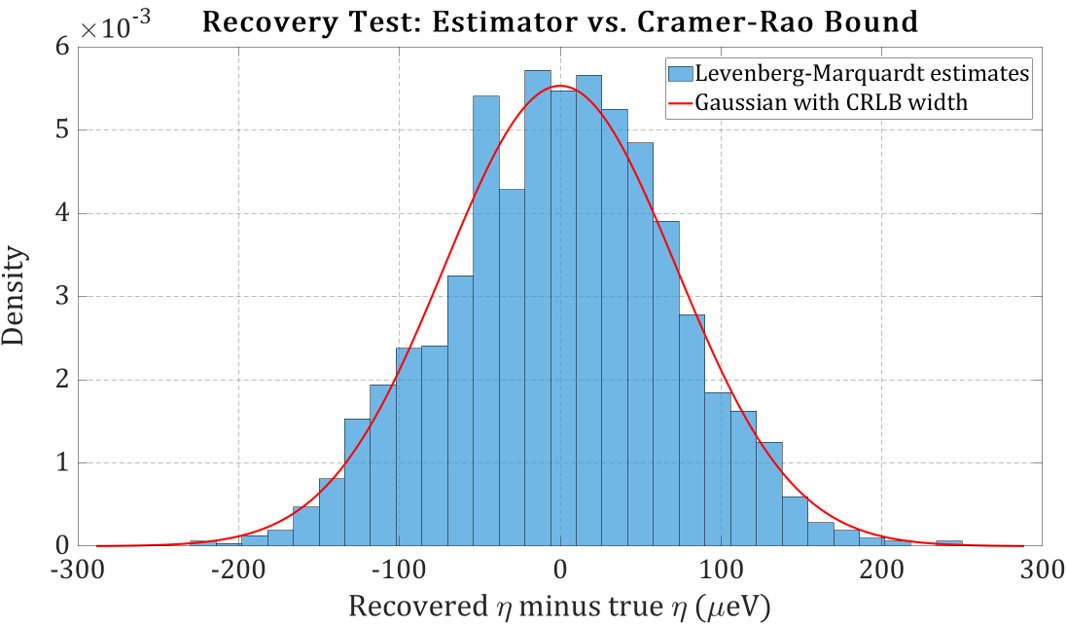}
    \caption{\label{fig:recovery} Distribution of the recovered $\eta$ minus its true value over 2,000 synthetic trials at $E = 100$ V/cm (blue histogram), and the Gaussian with the Cram\'er-Rao width (red curve).}
\end{figure}

In each of 2,000 trials, $M_0$ and $\eta$ are recovered with Levenberg-Marquardt updates from a starting point 5\% from $M_0$ and 70\% from $\eta$. The Cram\'er-Rao bound on $\eta$ at this field is 72.1 $\mu$eV. The standard deviation of the recovered $\eta$ is 70.1 $\mu$eV, which is 0.972 times the bound and consistent with unity at the sampling uncertainty of 0.016 for 2,000 trials. The bias of the recovered $\eta$ is 0.05\% of its value, and 95.9\% of the trials fall within $\pm 1.96$ bound widths of the true value (Fig. \ref{fig:recovery}). The standard deviation of the recovered $M_0$ is likewise 0.972 times its bound.

\bibliographystyle{apsrev4-2}
\bibliography{ref}

\end{document}